\documentclass[12pt]{article} 

\usepackage[utf8]{inputenc} 

\usepackage{geometry} 

\usepackage{graphicx,subcaption}

\usepackage{url}
\usepackage{pdflscape}
\usepackage{tabularx}
\usepackage{grffile}
\usepackage{booktabs} 
\usepackage{multirow}
\usepackage{array} 
\usepackage{paralist} 
\usepackage{verbatim} 
\usepackage{subfig} 

\usepackage{fancyhdr} 
\usepackage{bbm}
\usepackage{amsmath}
\usepackage{amsthm}
\usepackage{amssymb}
\usepackage{color}
\usepackage{url,algorithm,algorithmic}
\usepackage{paralist}

\theoremstyle{plain}
\newtheorem{thm}{Theorem}[section]

\newtheorem{prop}[thm]{Proposition}

\theoremstyle{definition}
\newtheorem{defn}{Definition}

\theoremstyle{remark}
\newtheorem{rem}{Remark}[section]

\newenvironment{pf}{{\noindent\sc Proof. }}{\qed\newline}

\def\R{{\mathbb R}}

\def\L{{\mathcal L}}

\def\E{{\mathbb E}}
\def\diag{{\rm diag}}

\def\Var{{\rm Var}}

\DeclareMathOperator{\argmax}{argmax}
\DeclareMathOperator{\argmin}{argmin}

\DeclareMathOperator{\rank}{rank}
\DeclareMathOperator{\tr}{tr}

\newcommand{\real}{\mathbb{R}}
\newcommand{\tran}{\mathsf{T}}
\newcommand{\dnorm}{\mathcal{N}}
\newcommand{\simiid}{\stackrel{\mathrm{iid}}\sim}
\newcommand{\err}{\mathrm{Err}}
\newcommand{\opt}{\mathrm{opt}} 

\newcommand{\phz}{\phantom{0}}

\title{Bi-cross-validation for  factor analysis}
\author{Art B. Owen \\Stanford University \and Jingshu Wang\\Stanford University}
\date{August 2015}

\begin{document}
\maketitle

\begin{abstract}
Factor analysis is over a century old, but
it is still problematic to choose the number
of factors for a given data set. 
We provide a systematic review of current methods and then 
introduce a method based on bi-cross-validation, using randomly
held-out submatrices of the data 
to choose the optimal number of factors.  
We find it performs better than 
many existing methods especially when both the 
number of variables and the sample size are large and
some of the factors are relatively weak.
Our performance criterion is based on recovery of an underlying
signal, equal to the product of the usual factor and loading matrices. 
Like previous comparisons, our work is simulation based.
Recent advances in random matrix theory provide principled choices for
the number of factors when the noise is homoscedastic, but not for
the heteroscedastic case.
The simulations we chose are designed using guidance from random matrix
theory. In particular, we include factors which are asymptotically 
too small to detect, factors large enough
to detect but not large enough to improve the estimate, and two classes of
factors (weak and strong) large enough to be useful. 
We also find that a form of early stopping regularization improves the
recovery of the signal matrix.
\end{abstract}

\section{Introduction}

Factor analysis is a core technology for handling large data matrices,
with applications in signal processing \cite{wax1985, hermus2007}, bioinformatics
\cite{price2006principal, patterson2006, leek2008general,  sun2012, gagnon2012using},
finance and econometrics \cite{forni2001, bai2008large}, and other areas
\cite{love2004factor, hochreiter2006new, kritchman2008determining}. 
In psychology, the factor model dates back at least to the paper of Spearman \cite{spearman1904} in 1904.
A basic factor analysis model assumes that the data 
matrix $Y\in\real^{N\times n}$ with $n$ observations and $N$ variables 
is represented as a matrix $X$ of some
low rank $k$ (the signal) plus independent heteroscedastic noise.  The signal $X$ in turn
can be factored into an $N\times k$ matrix times a $k\times n$ matrix
and this (nonunique) factorization may then be interpreted as a product
of latent variables times loading coefficients.

It is surprisingly difficult to choose the number $k$ of factors.
In traditional factor analysis problems which have a small $N$ but a relatively large $n$, 
there is no widely agreed best performing methods 
(see for example \cite{peres2005many})
and recommendations among
them are based largely on simulation studies  
\cite{cattell1977comprehensive, velicer2000construct}. 
Classical methods such as hypothesis testing 
based on likelihood ratios \cite{lawley1956tests} 
or methods based on information theoretic criteria \cite{wax1985}  
assume homoscedastic noise while heteroscedastic noise is more common in applications.
In addition, they are derived in an asymptotic
regime with a growing number of observations and fixed number
of variables and do not perform well on matrices where both dimensions are large.
Special methods for big data matrices where both $N$ and $n$ are large
have been proposed recently in the econometrics community 
\cite{bai2002determining, onatski2010, kapetanios2010, alessi2010, ahn2013}.
They are derived in an asymptotic framework
where the factor strength grows as
$N$ and $n$ both tend to infinity.
However, these methods may not work well on weaker factors
and that is a potential flaw when the strong factors are already well known
and we are trying to discover the weaker ones.
The random matrix theory (RMT) literature by contrast focuses on weak factors
but their methods are not well suited to heteroscedastic noise.  As a result, the present state
of theory does not provide usable guidelines.
This is a significant gap, because
the performance of factor analysis in many applications depends critically on 
the  number of factors chosen \cite{gagnon2012using, jolliffe2005principal}. 


In this paper, we develop an
approach to choosing the number of factors using bi-cross-validation (BCV)
\cite{owen2009}.
Our BCV involves holding out some rows
and some columns of $Y$, fitting a factor model to the held-in
data and comparing held-out data to corresponding fitted values.
We derive our method using recent insights from random matrix theory.
We test our method empirically using test cases that are also
designed using insights from RMT.
Our goal is not to recover the true number $k$ of factors, 
but instead to choose the number $k$ that lets us best recover the 
signal matrix $X$.  Using the true number of factors will
lead to a noisy estimate of $X$ when some factors are too weak 
to detect.

Based on previous theoretical results, we employ a taxonomy
dividing factors into four types based on their strength in an
asymptotic setting where both $n$ and $N$ go to infinity. 
To overcome identifiability problems, we assume that the factors 
are orthogonal to each other. Our factors may thus be linear combinations
of some real world factors.
The four factor levels are:
undetectable, harmful, helpful,  and strong. 

Strong factors are those that asymptotically explain a fixed percentage
of variance in the matrix $Y$. They become easy to detect as the 
corresponding singular values go to infinity under the asymptotics, but 
their presence causes difficulties for some methods of choosing $k$ 
when there are also weak factors.
The other factor types are weak
and explain a fraction of variance approaching some
limit $c/N$ as $n,N\to\infty$ with $N/n\to\gamma$.  
If $c$ is small compared to a detection threshold, then a singular value decomposition (SVD) 
based method can not distinguish 
that factor from noise, and the factor is undetectable.
If $c$ is somewhat larger, then that factor can be
detected but the corresponding eigenvectors cannot be estimated
accurately enough for that factor to improve estimation of $X$.
Such factors are harmful because detecting them can lead to worse performance.
If $c$ is still larger, then we can not only detect the factor but including
it in $\hat X$ yields an improvement. We call those factors helpful. 
Strong factors are also helpful, but `helpful' by itself will refer to helpful weak factors.
This taxonomy is based on homoscedastic Gaussian noise. 
A similar idea of the taxonomy also appeared in Onatski \cite{onatski2012, onatski2015}.
In \cite{onatski2012}, he proposed the model with both strong and weak factors, 
and an ``effective number'' of factors which is the number of 
detectable factors in our taxonomy. In \cite{onatski2015}, 
there is a concept of optimal loss efficiency which 
is attained by estimating the number of useful factors.

This paper is organized as follows. In section~\ref{sec:prob} we 
specify the factor model we study, the asymptotic regime, and our
estimation criterion.
Section~\ref{sec:prior}
reviews prior work on rank selection and determining the number of 
factors. 
It defines the boundaries in our four level taxonomy of factor sizes.
Section~\ref{sec:signalfromrank} describes our early stopping alternation (ESA)
algorithm to estimate the low-rank signal matrix with a given 
target $k$ for the number of factors. Section~\ref{sec:bicross}
introduces the BCV technique to determine the number of factors. 
Section~\ref{sec:simu} summarizes extensive simulation results.
In those cases
BCV is more reliably close to an oracle's performance than
all the other methods compared, including parallel analysis (PA), 
several leading methods in the econometrics literature and 
the information criteria based method \cite{nadakuditi2008sample} using RMT assuming white noise.
Also, unlike other methods, BCV becomes more likely to choose the unknown best
rank as sample size increases.
Section~\ref{sec:data} illustrates the BCV choice of $k$ on
some data sampled from a meteorite.
Section~\ref{sec:conclusion} concludes the paper.
An Appendix includes a detailed account of the simulations.

\section{Problem Formulation}\label{sec:prob}
Our data matrix is $Y \in \R^{N \times n}$ 
with a row for each variable and a column for each observation.
In the bioinformatics problems we have worked on, it is usual to have $N>n$ or even $N\gg n$, 
but this is not assumed.
In a factor model, $Y$ can be decomposed into a low rank signal matrix plus noise:
\begin{align}\label{model}
	Y &= X + \Sigma^{\frac{1}{2}} E  = LR + \Sigma^{\frac{1}{2}} E,
\end{align}
where the low rank signal matrix $X \in \R^{N \times n}$ is a product of factors 
$L \in \R^{N \times k_0}$ and $R \in \R^{k_0 \times n}$, both of rank $k_0$.
The noise 
matrix $E\in\real^{N\times n}$ has independent and identically distributed (IID)
entries with mean $0$ and variance $1$.
 Each variable has its own noise variance given by
$\Sigma = \diag (\sigma_1^2, \sigma_2^2, \cdots, \sigma_N^2)$.
The signal matrix $X$ is a signal that we wish to recover despite the
heteroscedastic noise.

The factor model is usually applied when we anticipate that
$k_0\ll\min(n,N)$.  Then identifying those factors suggests possible
data interpretations to guide further study.  When the factors correspond
to real world quantities there is no reason why they must be few in number
and then we should not insist on finding them all in our data 
as some factors maybe too small to estimate. 
We should instead seek the relatively important ones, 
which are the factors that are strong enough to
contribute most to the signals and be accurately estimated.

In a typical factor analysis, $R$ has $n$ IID columns corresponding to factors
and $L$ has nonrandom loadings.
We work conditionally on $R$ so that $X$  becomes a fixed
unknown matrix. A typical factor analysis aims to estimate the individual 
factors $L$ and $R$. To avoid identification problems, 
our goal is to recover $X$, seeking to minimize
\begin{align}\label{loss}
	\err_X(\hat X) \equiv \E\bigl(\|\hat X - X\|_F^2\bigr).
\end{align}
This criterion was used for factor models in \cite{onatski2015}
and for truncated SVDs and nonnegative matrix factorizations in \cite{owen2009}.
The estimate $\hat X$ can be factored into $\hat L$ and $\hat R$ using
rotations for greater interpretability.
  

\begin{defn}[Oracle rank and estimate]
Let $M$ be a method that for each integer $k\ge0$ gives a rank $k$ 
estimate $\hat X^M(k)$ of $X$ using $Y$ from model \eqref{model}. 
The oracle rank for $M$ is
\begin{align}\label{optk}
	k_M^* = \argmin_k \bigl(\|\hat X^M(k) - X\|_F^2\bigr),
\end{align}
and the corresponding oracle estimate of $X$ is
\begin{align}\label{goal}
	\hat X_{\text {opt}}^M = \hat X^M\bigl(k_M^*\bigr).
\end{align}
 \end{defn}

If all the factors are strong enough, then for a
good method $M$, we anticipate that $k_M^*$ should equal the true 
number of factors $k_0$. With weak enough factors we will have $k_M^*<k_0$.

Our algorithm has two steps.  First we need to devise a method
$M$ to effectively estimate $X$ given 
the oracle rank $k_M^*$.  Then with
such a method in hand, we need a means to estimate $k_M^*$.
Section~\ref{sec:signalfromrank} describes our early stopping
alternation (ESA) algorithm for estimating $X$ at a given $k$, 
which has the best performance compared with other methods given 
their own oracle ranks.
Then Section~\ref{sec:bicross} describes our BCV
for estimating $k_\mathrm{ESA}^\star$ for the ESA algorithm. First we describe
previous methods 
and the relevant RMT that motivates our comparisons.



\section{Literature review and factor taxonomy}\label{sec:prior}

Here we review the most commonly used methods for choosing
the number of factors. 
We begin with some classical methods in factor analysis 
which are typically based
on a limit with $n\to\infty$ while $N$ is fixed.
Then we consider some recently developed methods from the econometrics community for large matrices with 
strong factors and methods. The third source of methods are those
based on RMT which emphasizes weak factors with noise of constant variance. 
We use the recent work in RMT to develop the four level
taxonomy of factor sizes that guides our simulations.

\subsection{Classical methods for factor analysis}\label{sec:classic}
The most widely used classical methods for determining the number of factors 
or principal components include the 
scree test \cite{cattell1966scree, cattell1977comprehensive}, sphericity  
tests based on likelihood ratio \cite{bartlett1954note, lawley1956tests},  
parallel analysis (PA) \cite{horn1965rationale,
buja1992remarks}, the minimum average partial test of 
\cite{velicer1976determining} and information criteria based methods such as 
minimum description length (MDL) \cite{wax1985, fishler2002}. 
Those methods are aimed at estimating the true number $k$ of factors. 
They are derived for a setting where $n\to\infty$ with $N$ fixed. 
In that case, both the 
maximum-likelihood estimation of the factors and the sample covariance matrix will 
be consistent, 
thus $k_M^* = k_0$ asymptotically for a reasonable estimation method $M$. 

Regarding classical methods, we should mention the conceptual
difference between determining the number of principal components 
for principal component analysis (PCA) 
and determining the number of factors for factor analysis. 
Factor analysis has additive heteroscedastic noise 
that is not present in PCA. 
Though many of the above methods have been modified to be applied to both problems, 
 theoretical guarantees were only derived 
for PCA assuming white and Gaussian noise. 
Many researchers \cite{kaiser1960application,buja1992remarks,zwick1986comparison, velicer2000construct} 
have found out that those methods usually perform much better for estimating the principal components than for factor analysis.  
Some of them \cite{zwick1986comparison, velicer2000construct} suggest that even for factor analysis, 
one should perform PCA  
first in the initial stage to determine the number of factors before estimating the factors. 
We adopt this suggestion in this paper later when comparing these methods 
in Section~\ref{sec:simu}.

There is a large amount of evidence 
\cite{zwick1986comparison, hubbard1987empirical, velicer2000construct, peres2005many} 
that PA is one of the most accurate of the above classical methods 
for determining the number of factors. 
Parallel analysis compares the observed eigenvalues of the correlation matrix 
to those obtained 
in a Monte Carlo simulation.  The first factor is retained if and only 
if its associated eigenvalue is larger than the $95$'th percentile of 
simulated first eigenvalues. For $k\ge2$,
the $k$'th factor is retained when the first $k-1$ factors were retained 
and the observed $k$'th 
eigenvalue is larger than the $95$'th percentile of simulated $k$'th factors. 
The permutation version of PA was introduced by 
\cite{buja1992remarks}.  There the eigenvalues are simulated 
by applying independent uniform random permutations to 
each of the variables stored in $Y$. 
The earlier method of Horn \cite{horn1965rationale} resamples from a Gaussian
distribution.  
Parallel analysis has been used recently in bioinformatics 
\cite{leek2008general, sun2012}. Though there exist no theoretical results to 
guarantee the accuracy of PA, it performs very well in practice. 


\subsection{Methods for large matrices and strong factors}\label{sec:strong-factor}

This collection of methods is designed for an asymptotic regime
where both $n, N\to \infty$ while $k$ is fixed. 
For strong factors, it is usually assumed that $RR^T/n \to \Sigma_R$ and 
$L^TL/ N \to \Sigma_L$ for some $k_0 \times k_0$ positive definite matrices 
$\Sigma_R$ and $\Sigma_L$. In that case, the singular values of 
$X$ are $O(\sqrt {nN})$. The methods are designed to estimate the true number 
of factors. In the above framework, the factors can be estimated consistently, and we should expect 
$k_M^* = k_0$. This was proved when $M$ is the SVD by Onatski \cite{onatski2015}. 

Some of the most popular methods to estimate the number of factors 
under the above scenario are based on the information criteria developed by 
Bai and Ng \cite{bai2002determining}, with later improvements in \cite{alessi2010}. 
It has been shown that these information criteria based 
rules are asymptotically consistent. Kapetanios \cite{kapetanios2004, kapetanios2010} 
proposed several methods assuming strong factors but making use of the RMT results on the 
sample eigenvalue distribution of pure white noise. However, the theoretical guarantees for
his methods require homescedastic noise. Ahn and Horenstein \cite{ahn2013} 
recently proposed two estimators for determining the number of factors by simply 
maximizing the ratio of two adjacent eigenvalues of sample covariance matrix. The idea of 
maximizing such a ratio to estimate the number of factors 
can be also found in \cite{lam2012, lan2014}. 
Onatski \cite{onatski2010} developed an estimator (ED) based on the 
difference of two adjacent eigenvalues of sample covariance matrix, and has proved its 
consistency under a weaker assumption of the factor strength: instead of growing in the order of $O(\sqrt N)$, 
the singular values of $X/\sqrt n$ are just required to diverge in probability as $N \to \infty$.
For econometrics applications, there are more methods to estimate the number of factors 
\cite{forni2000, amengual2007, hallin2007} for dynamic factor models.
These models assume a times series structure on the factors. Such dependency models
are beyond the scope of this paper.


\subsection{Methods for large matrices and weak factors}\label{sec:rmt}
Here we review methods to estimate the number of weak factors in white noise, based on 
results in RMT. In this asymptotic regime, $n$ and $N$ diverge to infinity
while $k$ is fixed and the singular values of $X$ are $O(1)$.
The model is commonly framed as
\begin{align}\label{white}
	Y = \sqrt{n}UDV^\tran + \sigma E
\end{align}
where $\sqrt n UDV^\tran$ is the SVD of $X$, so that $U\in\real^{N\times k_0}$ and $V\in\real^{N \times k_0}$ 
satisfy $U^\tran U = V^\tran V = I_{k_0 \times k_0}$. The matrix $D = \diag(d_1, d_2, \cdots, d_{k_0})$ defines the strength of each signal. 
Asymptotically,  $d_i^2 \to u_i$ for some constants $u_i$.
The noise matrix $E\in\real^{N\times n}$ is usually taken to have 
IID entries with mean $0$, variance $1$ and finite fourth moment \cite{baik2006}.

Estimation of $X$ is typically through the singular value decomposition (SVD)
of $Y$, retaining the fitted singular vectors, but shrinking or
truncating the corresponding singular values.
In the limit $n, N \to \infty$ and ${N}/{n} \to \gamma$, there is a well known phase transition 
for signal detection.
If  $u_i^2 < \sigma^2\sqrt{\gamma}$ then the corresponding
factor is asymptotically not detectable using SVD based methods,
while if $u_i^2>\sigma^2\sqrt{\gamma}$ the factor can be detected.
See \cite{paul2007asymptotics, benaych2012singular, 2009arXiv0909.3052P}
for statements of this result. Simulations \cite{nadakuditi2008sample, gavish2014} 
have also confirmed this result.

A principled way to select the rank is to estimate the number of factors with $u_i$ above 
the asymptotic detection threshold $\sigma^2\sqrt{\gamma}$.
Nadakuditi and Edelman \cite{nadakuditi2008sample} used an 
information criteria based method modified 
from the classical MDL estimator \cite{wax1985}. 
Kritchman and Nadler \cite{kritchman2008determining} 
developed an algorithm based on a sequence of hypothesis tests 
which are connected with the Roy's classical largest root test \cite{roy1953} to check for 
sphericity of a covariance matrix. Both methods will consistently estimate
the number of detectable factors under weak factor asymptotics. 
Similar to \cite{kritchman2008determining}, 
\cite{choi2014} provides a sequential hypotheses testing method 
which is not based on asymptotics.

Neither the true rank, nor the number of detectable factors will necessarily optimize
our criterion \eqref{optk}. 
The problem is that a factor stronger than the detection threshold
might still not be strong enough to allow adequate estimation of the
corresponding singular vectors.
Owen and Perry \cite{owen2009} propose a BCV
algorithm to choose $k$ for the truncated SVD, motivated by the loss~\eqref{loss}.
Perry's work \cite{2009arXiv0909.3052P}  on BCV
identifies a higher threshold for $u_i$ beyond which including
the corresponding singular vectors reduces the loss~\eqref{loss}.
He also shows that the rank selected by BCV will
track the oracle's rank for truncated SVD; his formal statement
is in Theorem~\ref{bcv_thm} below. 
This second estimation threshold was later derived by
\cite{gavish2014} and by~\cite{onatski2015}. 

The above results are only valid in the white noise model (\ref{white}), which is much 
simpler than the heteroscedastic model (\ref{model}). For 
more general noise covariance structures, 
there are several recent theoretical results, 
but none of them solve our problem.
For example, Nadler \cite{nadler2008finite} considered 
a general spiked covariance model 
with the eigenvalues corresponding to the noise in the population covariance 
matrix converging to some limiting distribution.
However, our heteroscedastic model (\ref{model}) is not directly related to a 
spiked covariance model. Nadakuditi \cite{nadakuditi2014optshrink} developed a
 method to shrink singular values to recover a low-rank signal matrix with noise 
from a class of distributions more general than IID Gaussian. But he assumed that 
either the noise matrix or the signal matrix is bi-orthogonally invariant, and 
he did not show how to estimate the rank. Onatski \cite{onatski2012} considered noise 
whose covariance structure can be represented by a Kronecker product, which includes the 
heteroscedastic noise case. However, his theory depends on the strong assumption that 
the factors and the noise covariance have the same eigenvectors. He suggested 
using the ED estimator mentioned in Section~\ref{sec:strong-factor} to estimate 
the number of weak detectable factors, which works well in his simulations.

\subsection{Factor categories and test cases}\label{sec:testcases}

When we simulate the factor model for our tests, 
we will generate it as 
\begin{align}\label{eq:genvia}
Y=\Sigma^{1/2}( \Sigma^{-1/2}X+E)=\Sigma^{1/2}( \sqrt{n}UDV^\tran+E).
\end{align}
The matrix $\Sigma^{-1/2}X=\sqrt{n}UDV^\tran$ has the same low rank that $X$
does. Here $UDV^\tran$ is an SVD and we generate 
the matrices $U$ and $V$ from appropriate distributions. 
The normalization in~\eqref{eq:genvia} allows us to make direct use of RMT
in choosing $D$.  The matrix $V$ is uniformly distributed, but $U$ has
a non-uniform distribution to avoid making rows with large mean squared $U$-values
coincide with rows having large $\sigma_i$. Such a coincidence could make
the problem artificially easy.
See the Appendix for a description of the sampler.

Based on the discussion in Section~\ref{sec:rmt} and under the asymptotics that 
$n, p \to \infty$, we may place each
factor into a category depending on the size of $d_i^2$.
The categories are:
\begin{enumerate}
\item Undetectable: $d_i^2$ is below the detection threshold, thus the factor is 
  asymptotically undetectable by SVD based methods.
\item Harmful: $d_i^2$ is above the detection threshold but 
below the threshold at which their inclusion in the model improves accuracy.
\item Useful: $d_i^2$ is above the detection threshold
but is $O(1)$. It contributes an $N\times n$ matrix to $Y$
with sum of squares $O(n)$, while the expected sum of squared errors
is $nN\sigma^2$.
\item Strong: $d_i^2$ grows proportionally to $N$. The factor sum of
squares is then proportional to the noise level.  
\end{enumerate}

Undetectable factors essentially add to the noise level.
Asymptotically, no method based on sample eigenvalues can detect them, and so
they play a small role in determining which
method to choose $k$ is best.

Harmful factors can cause severe difficulties for 
a factor number estimator to reduce the loss~\eqref{loss}. 
They are large enough to be detected but
including them makes the loss~\eqref{loss} larger. Changing
an algorithm to better detect such factors could lead it to
have worse performance.

Useful weak factors are large enough that including them
reduces the loss.  It is generally not possible to estimate
their corresponding eigenvectors consistently. The estimated
and true eigenvectors only converge in a limit where $d_i^2$
is an arbitrarily large constant. Separating useful from harmful
weak factors is important for accurate estimation of $X$.

The strong factors are large enough to be almost unmissable.
When one or more of them is present they may very well
put a clear knee in the scree plot, though that knee won't 
necessarily be at the optimal $k$ when there are also some useful weak factors.
Given an estimation method, 
the total number of useful weak factors and strong factors is 
the same as the oracle rank.

Real data often have include factors that fit
the asymptotic strong factor category. 
In a matrix of dimensional measurements on animals, there is likely
to be a strong factor for the overall size of those animals.
In educational testing data where $n$ students each answer $N$
questions there is very often a strong factor interpreted as student
ability with a corresponding loading for item difficulty.
In modeling daily returns of stocks there may be one factor corresponding
to overall market movements that affect all stocks.
Although strong factors should be easy to detect,
they can cause severe difficulties for some algorithms as illustrated 
in Section~\ref{sec:simu}. Useful
weak factors may appear negligible in comparison to the strong ones.
In each of these examples one can envision settings where the 
strongest factors are obvious and uninteresting while the
weak factors have useful insights.

Strong factors resemble the giant components commonly found
in networks \cite{easley2010networks}. 
Network theory has several well understood
mechanisms which lead to giant components.
A mechanism for strong versus weak factors seems to be missing.
Suppose that  one keeps adding measurements, increasing $N$, and 
perhaps doing so by adjoining additional features that are less and
less important to one's primary scientific goals.  A factor that strongly
predicts the first few variables but is only weakly related to subsequent
ones might become a weak factor  in such a limit.  A factor related to
all of the variables we add would ordinarily be a strong one.

In the following sections we compare methods using the six
testing scenarios described in Table~\ref{factor_strength_table}.
They have been customized based on our goals and our understanding
of the problem.
All of these cases have eight nonzero factors of which one is undetectable.
We anticipate that the number of harmful factors is an important
variable, and so it generally increases with scenario number,
ranging from $1$ to $6$.  The remaining factors are split between
strong and merely useful. By including several scenarios with equal numbers of
harmful factors, we can vary the ratio of strong to useful factors at high
and low numbers of harmful factors.

In the white noise model, the category that a factor falls into
depends on the ratio $d_i^2/(\sigma^2\sqrt{\gamma})$.
When we simulate factors we use the same critical ratios
but replace $\sigma^2$ by $(1/N)\sum_{i=1}^N\sigma^2_i$.

\begin{table}
\centering 
\begin{tabular}{@{}lrrrrrr@{}}
  \toprule 
  & \multicolumn{6}{c}{Scenario}\\
  \cmidrule(lr){2-7}
  & 1 & 2 & 3 & 4 & 5 & 6\\
    \cmidrule(lr){1-7}
\# Undetectable & 1 & 1 & 1 & 1 & 1 & 1\\
\# Harmful &1 & 1 & 1 & 3 & 3 & 6\\
\# Useful  & 6 & 4 & 3 & 1 & 3 & 1\\
\# Strong & 0 & 2 & 3 & 3 & 1 & 0\\
\bottomrule 
\end{tabular}
\caption{Six factor strength scenarios considered in our simulations. 
}
\label{factor_strength_table}
\end{table}

For each of these six cases
we consider various levels of noise variance.
The $\sigma^2_i$ are independent inverse gamma random
variables with mean $1$ and variances $0$ or $1$ or $10$.
We also consider $5$ aspect ratios,
$N/n\in \{0.02, 0.2, 1, 5, 20\}$.
For each aspect ratio we consider two sizes $n$.
That is, we consider $6\times 3\times 5\times 2=180$
cases spanning a wide range of problems.
The complete details are in the Appendix.

\section{Estimating $X$ given the rank $k$}\label{sec:signalfromrank}

Here we consider how to estimate $X$
using exactly $k$ factors.
This will be the inner loop for an algorithm that tries various $k$. 
The goal in this section 
is to find a method that has good performance when given its oracle rank.
Assuming Gaussian noise, we get the log-likehihood function:
\begin{align}\label{log-likelihood}
	\log \L(X, \Sigma)  = -\frac{Nn}{2}\log(2\pi) - \frac{n}{2}\log \det \Sigma +
 \tr \Bigl[-\frac{1}{2}\Sigma^{-1}(Y - X)(Y - X)^\tran\Bigr].
\end{align}
If $\Sigma$ were known it would be straightforward to estimate $X$ 
using an SVD,
but $\Sigma$ is unknown.  Given an estimate of $X$ it is
straightforward to optimize the likelihood over $\Sigma$.  
Next we describe our
alternating algorithm and we employ an early stopping rule to
regularize it.


The truncated SVD of a matrix $Y$ is
\begin{align}\label{trunc-svd}
	Y(k) = U(k)D(k)V(k)^\tran
\end{align}
where $D(k)$ is the diagonal matrix of the $k$ largest singular values of $Y$, and 
$U(k)$ and $V(k)$ are the matrices of the corresponding singular vectors. 
We start with an initial estimate of $\Sigma$ using the sample variance: 
\begin{align}\label{initial_Sigma}
	\hat\Sigma = \diag\Bigl(\Bigl(Y - \frac{1}{n}Y\mathbf{1}_{n \times n}\Bigr)
\Bigl(Y - \frac{1}{n}Y\mathbf{1}_{n \times n}\Bigr)^\tran\,\Bigr).
\end{align}
Given an estimate $\hat\Sigma$, our rank $k$ estimate $\hat X$ is
the truncated SVD of the reweighted matrix 
$\tilde Y = \hat\Sigma^{-\frac{1}{2}}Y$:
	\begin{align}\label{eq:getxhat}
		\hat X = \hat \Sigma^{\frac{1}{2}}\tilde Y(k).
	\end{align} 
Given an estimate $\hat X$, our new variance estimate $\hat\Sigma$ contains the mean squares of the 
residuals: 
\begin{align}\label{eq:getsigmahat}
\hat{\Sigma} = \frac{1}{n} \diag \bigl[\bigl(Y - \hat X\bigr)\bigl(Y - 
\hat X\bigr)^\tran\bigr].
\end{align}

Both of the above two steps can increase $\log \L(X, \Sigma)$ but not decrease it.
Simply alternating those two steps to convergence is not effective.
The algorithm often does not converge.  Nor should it, because the likelihood
is unbounded as even one of the $\sigma_i$ decreases to zero.
Such a degenerate problem is similar to 
the degenerate problem 
when one tries to fit real valued
data to a mixture of two Gaussians. In that case the likelihood is unbounded
as one of the mixture components converges to a point mass 
(the variance of one component goes to $0$).


It is not straightforward to prevent $\sigma_i$ from approaching $0$.
Imposing a bound $\sigma_i\ge\epsilon>0$ leads to some
$\sigma_i$ converging to $\epsilon$.
There are numerous approaches to regularizing $\hat X$ to prevent
$\sigma_i\to0$.
One could model the $\sigma_i$ as IID from some prior distribution.
However, such a distribution must also avoid putting too much mass near
zero.  We believe that this transfers the singularity avoidance problem
to the choice of hyperparameters in the $\sigma$ distribution and does
not really solve it. We have also found in trying it that even when
$\sigma_i$ are really drawn from our prior, the algorithm still 
converged towards some zero estimates.

A second, related approach is to employ a penalized likelihood
\begin{align}\label{penalizedLoss}
	L_\text{reg}\bigl(Y, \lambda, \hat X, \hat \Sigma\bigr) = -n\log \det {\hat\Sigma} + \tr \bigl[\Sigma^{-1}(Y - \hat X)(Y -\hat X)^\tran\bigr] + \lambda P\bigl(\hat \Sigma\bigr),
\end{align}
where $P$ penalizes small components $\sigma_i$.
This approach has two challenges.  It is hard to select a penalty 
$P$ that is strong enough
to ensure  boundedness of the likelihood, 
without introducing too much bias. 
Additionally, it requires a choice of $\lambda$. Tuning $\lambda$ by
cross-validation within our bi-cross-validation algorithm is unattractive. Also there is
a risk that cross-validation might choose $\lambda=0$ allowing one or
more $\sigma_i\to0$.

We do not claim that these methods cannot in the future
be made to work. They are however not easy to use, and we found
a simpler approach that works surprisingly well.
Our approach is to employ early stopping. We start at \eqref{initial_Sigma}
and iterate the pair~\eqref{eq:getxhat} and~\eqref{eq:getsigmahat}
some number $m$ of times and then stop. 

To choose $m$,
we investigated $180$ test cases based
on the six factor designs in Table~\ref{factor_strength_table}, 
three dispersion levels for the $\sigma^2_i$, five aspect
ratios $\gamma$ and $2$ data sizes. The details are in the Appendix.
The finding is that taking $m=3$ works almost as well as if we
used whichever $m$ gave the smallest error for each given data set.  

More specifically, define the oracle estimating error 
using early stopping at $m$ steps as
\begin{align}\label{errX}
\err_X(m) =\min_k\|\hat X^m(k) - X\|^2_F 
\end{align}
where $\hat X^m(k)$ is the estimate of $X$ using $m$ iterations and rank $k$.
We judge each number $m$ of steps,  by the best $k$ that might be used with it.

For early stopping alternation (ESA),
we define the oracle stopping number of steps on a data set as
\begin{align}\label{optm}
m_{\opt} = \argmin_m\err_X(m)  = \argmin_m\min_k\|\hat X^m(k)-X\|^2_F.
\end{align}
We have found that $m = 3$ is very nearly
optimal in almost all cases. 
We find that $\err_X(3)/\err_X(m_\opt)$
is on average less than $1.01$, with a standard deviation of $0.01$ (see Appendix). 
Using $m=3$ steps with the best $k$
is nearly as good as using the best possible combination of $m$ and $k$. 
We have tested early stopping on other data sizes, factor strengths and noise distributions, 
and find that $m=3$ is a robust choice.
Early stopping is also much faster than
iterating until a convergence criterion has been met.

In the Appendix, we compare  ESA to other methods for estimating $X$, 
including SVD, PCA (SVD after data standardization) and  
the quasi maximum-likelihood method (QMLE). 
The QMLE is derived by a classical factor analysis approach and 
it gives consistent estimation for strong factors and large datasets \cite{bai2012}. 
For the heteroscedastic noise cases and 
given the oracle rank of each method, 
ESA performs better than SVD and 
PCA in most cases. It also performs better than QMLE on average and when 
the aspect ratio $N/n$ is not too small. Comparing ESA with 
an oracle SVD method that knows the noise variance, we find that they have comparable 
performance. 

Given the above findings, we turn our attention to estimating the oracle $k$ 
for ESA in Section~\ref{sec:bicross}.

\begin{rem}
Early-stopping of iterative algorithms is a well-known regularization strategy for inverse problems and training machine learning models like neural networks and boosting \cite{yao2007early, zhang2005boosting, hastie2009elements, caruana2001overfitting}. An equivalence between early-stopping and adding a penalty term 
has been demonstrated in some settings
\cite{fleming1990equivalence, rosset2004boosting}. 
\end{rem}

\begin{rem}
ESA starting from \eqref{initial_Sigma} with $m = 1$ is equivalent to 
PCA. 
Using $m>1$ iterations can be interpreted as using an estimated
signal matrix to improve the estimation of $\Sigma$, so
ESA with $m = 3$ can be understood as applying truncated SVD on a more properly reweighted data than one gets with $m=1$.
\end{rem}

\section{Bi-cross-validatory choice of  $k$}\label{sec:bicross}

Here we describe how BCV works in the
heteroscedastic noise setting.  Then we give our choice
for the shape and size of the held-out submatrix
using theory from \cite{2009arXiv0909.3052P}.
 
\subsection{Bi-cross-validation to estimate $k_\text{ESA}^*$}

We want $k$ to minimize  the squared estimating error \eqref{optk}
in $\hat X$.
We adapt the BCV technique of Owen and Perry \cite{owen2009}
to this setting of unequal variances.
We randomly select $n_0$ columns and $N_0$ rows as the held-out block and partition the data matrix $Y$ (by permuting the rows and columns) into four folds,
$$ Y = \left( 
		\begin{matrix}
			Y_{00} & Y_{01} \\
			Y_{10} & Y_{11}
		\end{matrix}
	\right) $$
	where $Y_{00}$ is the selected $N_0\times n_0$ held-out block, and the other three blocks $Y_{01}, Y_{10}$ and $Y_{11}$ are held-in. Correspondingly, we partition $X$ and $\Sigma$ as
$$ X = \left( 
		\begin{matrix}
			X_{00} & X_{01} \\
			X_{10} & X_{11}
		\end{matrix}
	\right),\quad\text{and} \quad
	\Sigma = \left( 
		\begin{matrix}
			\Sigma_{0} & 0 \\
			0 & \Sigma_{1}
		\end{matrix}
	\right).$$	
	The idea is to use the three held-in blocks to estimate $X_{00}$ for each candidate rank $k$ and then select the best $k$ based on the BCV estimated prediction error. 
	
We rewrite the model (\ref{model}) in terms of the four blocks:
\begin{align*}
	\left( \begin{matrix}
			Y_{00} & Y_{01} \\
			Y_{10} & Y_{11} \end{matrix}\right) 
	&= \left( \begin{matrix}
			X_{00} & X_{01} \\
			X_{10} & X_{11} \end{matrix}\right) + 
			\left( \begin{matrix}\Sigma_0 & 0\\0 & \Sigma_1\end{matrix}\right)^{\frac{1}{2}}
			\left( \begin{matrix}
			E_{00} & E_{01} \\
			E_{10} & E_{11} \end{matrix}\right) \\
	&=  \left( \begin{matrix}
			L_0 R_0 & L_0 R_1 \\
			L_1 R_0 & L_1 R_1 \end{matrix}\right) + 
			\left( \begin{matrix}
			\Sigma_0^{\frac{1}{2}}E_{00} & \Sigma_0^{\frac{1}{2}}E_{01} \\
			\Sigma_1^{\frac{1}{2}}E_{10} & \Sigma_1^{\frac{1}{2}}E_{11} \end{matrix}\right)
\end{align*}
where $L = \left( 
		\begin{matrix}
			L_{0} \\ L_{1}
		\end{matrix} \right)$
and $R = \left( 
		\begin{matrix}
			R_{0} & R_{1}
		\end{matrix} \right)$ are decompositions of the factors.

The held-in block 
$$Y_{11} = X_{11} + \Sigma_1^{\frac{1}{2}}E_{11}= L_1R_1 + \Sigma_1^{\frac{1}{2}}E_{11}$$ 
has the low-rank plus noise form, so we can use ESA to get estimates $\hat X_{11}(k)$ and $\hat \Sigma_1$ for a given rank $k$. Next for $k < \rank(Y_{11})$ we choose  rank $k$ matrices $\hat L_1$ and $\hat R_1$ with
\begin{align}\label{decomp}
	\hat X_{11}(k) = \hat L_1 \hat R_1.
\end{align}
Then we can 
estimate $L_0$ by solving $N_0$ linear regression models
$Y_{01}^\tran = \hat R_1^\tran L_0^\tran + E_{01}^\tran\Sigma_0^{1/2}$, 
and estimate $R_0$ by solving $n_0$ weighted linear regression models
$Y_{10} = \hat L_1 R_0 + \hat \Sigma_1^{1/2}E_{10}$.
These least square solutions are
$$\hat {R}_0 = (\hat L_1^\tran\hat\Sigma_1^{-1}\hat L_1)^{-1}
						\hat L_1^\tran\hat\Sigma_1^{-1}Y_{10},
\quad\text{and}\quad
\hat{L}_0 = Y_{01}\hat R_1^\tran(\hat R_1 \hat R_1^\tran)^{-1} $$
which do not depend on the unknown $\Sigma_0$. We get a rank $k$ estimate of $X_{00}$ as
\begin{align}\label{eq:x00hat}
\hat X_{00}(k) = \hat L_0 \hat R_0.
\end{align}

Though the decomposition (\ref{decomp}) is not unique, the estimate
$\hat X_{00}(k)$ is unique.
To prove it we need a reverse order theorem for Moore-Penrose
inverses.
For a matrix $Z\in\real^{n\times d}$, the
Moore-Penrose pseudo-inverse of $Z$ is denoted $Z^+$.
\begin{thm} \label{thm:macduffee}
	Suppose that $X = LR$, where $L \in \R^{m \times r}$ and $R \in \R^{r \times n}$ both have rank $r$. Then 
$X^+ = R^+L^+ = R^\tran(RR^\tran)^{-1}(L^\tran L)^{-1}L^\tran$. 
\end{thm}
\begin{pf}
This is MacDuffee's theorem. 
There is a proof in \cite{owen2009}.
\end{pf}

\begin{prop}
The estimate $\hat X_{00}(k)$ from~\eqref{eq:x00hat}
does not depend on the decomposition of $\hat X_{11}(k)$  in \eqref{decomp} and has the form
	\begin{align}\label{new_BCV}
	\hat X_{00}(k) = Y_{01}\bigl(\hat\Sigma_1^{-\frac{1}{2}}\hat X_{11}(k)\bigr)^+
	\hat\Sigma_1^{-\frac{1}{2}}Y_{10}.
\end{align}
\end{prop}
\begin{pf}
Let $\hat X_{11}(k) = \hat L_1\hat R_1$ be any decomposition
satisfying~\eqref{decomp}. Then
\begin{align}\label{inverse1}
\hat X_{00} &= \hat{L}_0\hat{R}_0\\ 
                &= Y_{01}\hat R_1^\tran\bigl(\hat R_1 \hat R_1^\tran\bigr)^{-1}
\bigl(\hat L_1^\tran \hat\Sigma_1^{-1}\hat L_1\bigr)^{-1}\hat L_1^\tran \hat\Sigma_1^{-1}Y_{10}\notag\\ 
	&= Y_{01}\bigl(\hat\Sigma_1^{-\frac{1}{2}}\hat L_1\hat R_1\bigr)^+\hat\Sigma_1^{-\frac{1}{2}}Y_{10}\notag
	= Y_{01}\bigl(\hat\Sigma_1^{-\frac{1}{2}}\hat X_{11}(k)\bigr)^+\hat\Sigma_1^{-\frac{1}{2}}Y_{10}.\notag\qedhere
\end{align}
The third equality follows from Theorem~\ref{thm:macduffee}.
\end{pf}

Next, we define the cross-validation prediction average squared error for block $Y_{00}$ as
\begin{align*}
	\widehat {\text{PE}}_k(Y_{00}) = \frac{1}{n_0N_0}\bigl\|Y_{00} - \hat X_{00}(k)\bigr\|_F^2.
\end{align*}
Notice that as the partition is random, we have:
\begin{align*}
	\E\bigl(\widehat {\text{PE}}_k(Y_{00})\bigr) = \E\left\{\frac{1}{n_0N_0}\text{Err}_{X_{00}}\bigl(\hat X_{00}(k)\bigr)\right\} + \frac{1}{N}\sum_{i=1}^N \sigma_i^2,
\end{align*}
where $\text{Err}_X(\hat X)$ is the loss defined at \eqref{loss}.
The expectation is over the 
noise and the random partition, for a fixed signal matrix.

The above random partitioning step is repeated independently $R$ times, 
yielding the average BCV mean squared prediction error for $Y$,
\begin{align*}
\widehat {\text{PE}}(k) = \frac{1}{R}\sum_{r = 1}^{R}\widehat {\text{PE}}_k(Y_{00}^{(r)}) .
\end{align*}
The BCV estimate of $k$ is then
\begin{align}\label{bcv_k}
	\hat k^* = \argmin_k \widehat {\text{PE}}(k).
\end{align}

We investigate integer values of $k$ from $0$
to some maximum. We cannot take
$k$ as large as $\min(n_1, N_1)$ where $n_1 = n - n_0$ and 
$N_1 = N - N_0$, for then we will surely get $\sigma_i=0$ even with early stopping.
We impose an additional constraint on $k$ to keep the diagonal of $\hat\Sigma_1$
away from zero. If for some $k$ we observe that
\begin{align}\label{eq:geomean}
	\frac{1}{N_1}\sum_{i=1}^{N_1} \log_{10}\bigl(|\hat\sigma_{i, 1}^{(k)}|\bigr) < - 6 + \log_{10}\bigl(\max_i|\hat\sigma_{i, 1}^{(k)}|\bigr)
\end{align}
where $\hat\Sigma_1(k) = \diag\bigl(\hat\sigma_{1, 1}^{(k)}, \hat\sigma_{2, 1}^{(k)}, \cdots, \hat\sigma_{N_1, 1}^{(k)}\bigr)$, then we do not consider
any larger values of $k$.
The condition~\eqref{eq:geomean} means that the geometric mean of the 
variance estimates is below $10^{-6}$ times the largest one.

\begin{rem}
Owen and Perry \cite{owen2009} 
mentioned that BCV can miss large but very sparse components in the SVD in a white noise model, and they suggested rotating the data matrix as a remedy. However, in our problem where the noise is heteroscedastic, there will be an identifiability issue between factors and noise if the factors are too sparse and the support of the low rank matrix is concentrated in a few locations (see for example \cite{chandrasekaran2010latent}). Thus, we only investigate cases where the signal matrix $X$ is not sparse, and do not use rotation to remove sparseness.
\end{rem}

\subsection{Choosing the size of the holdout $Y_{00}$}
We define the true prediction error for ESA as:
\begin{align*}
	\text{PE}(k) = \frac{1}{nN}\bigl\|X - \hat X(k)\bigr\|_F^2 + \frac{1}{N}\sum_i\sigma_i^2
\end{align*}
and then  the oracle rank is
$k_{\text{ESA}}^* = \argmin_k \text{PE}(k)$.

Ideally, we would like $\widehat {\text{PE}}(k)$ be a good estimate of $\text{PE}(k)$. 
For good estimation of $X$ it suffices
to have $\hat k^*$ (defined in (\ref{bcv_k})) be a good estimate of 
$k_{\text{ESA}}^*$. 

When it is known that $\Sigma = \sigma^2I$, we can use the truncated SVD to estimate $X$ and for BCV the estimate of $X_{00}$ simplifies to
\begin{align} \label{white_bcv_k}
	\hat X_{00}(k) = Y_{01}\bigl( Y_{11}(k)\bigr)^+Y_{10},
\end{align}
where $Y_{11}(k)$ is the truncated SVD in \eqref{trunc-svd}. 
Perry \cite{2009arXiv0909.3052P} 
proved that $\hat k^*$ and $k_{\text{ESA}}^*$ track each other asymptotically if the relative size of the held-out matrix $Y_{00}$ satisfies the following theorem.

\begin{thm} \label{bcv_thm}
For model (\ref{white}), if $k_0$ is fixed and ${N}/{n}\rightarrow \gamma\in(0,\infty)$ as $n \rightarrow \infty$, then 
$\hat k^*$ and $\argmin_k\E(\widehat{\mathrm{PE}}_k(Y_{00}))$ converge to the same value if 
\begin{align} \label{rho_chosen}
	\sqrt \rho = \frac{\sqrt 2}{\sqrt {\bar\gamma} + \sqrt{\bar \gamma + 3}}
\end{align} 
holds, where 
\begin{align*}
	\bar \gamma = \left(\frac{\gamma^{1/2} + \gamma^{-1/2}}{2}\right)^2,
\quad\text{and}\quad
	\rho = \frac{n - n_0}{n}\cdot\frac{N - N_0}{N}.
\end{align*}
\end{thm}

Here $\rho$ is the fraction of entries from $Y$ 
in the held-in block $Y_{11}$. 
The larger $\bar\gamma$ is, the smaller $\rho$ will be, thus $\rho$ reaches its maximum when $Y$ is square with $\gamma = 1$. 
For example, when $\gamma = 1$, then $\rho \approx 22\%$.  
In contrast, if $\gamma = 50 \text{ or } 0.02$, $\rho$ then drops to only $3.5\%$. 

Theorem~\ref{bcv_thm} compares the best $k$ for $\E(\widehat{\mathrm{PE}}_k)$ to
the best $k$ for the true error. 
Owen and Perry \cite{owen2009}  found that the BCV curve under repeated subsampling was
remarkably stable for large matrices, and then the best rank per sample
will be close to the one that is best on average.

In our simulations, we use \eqref{rho_chosen} to determine the size of $Y_{00}$. 
Further, to determine $n_0$ and $N_0$ individually,  we make $Y_{11}$ as square 
as possible as long as $n_0 \geq 1$ and $N_0 \geq 1$. For instance, with 
$\gamma=1$ as $\rho \approx 22\%$, 
we hold out roughly half the rows and columns of the data.

\section{Simulation Results}\label{sec:simu}

We use simulation scenarios described in 
Section~\ref{sec:testcases} and the Appendix. 
Those simulations have $\E(\sigma^2_i)=1$
but fall into three different groups:
white noise with $\Var(\sigma^2_i)=0$,
mild heteroscedasticity with $\Var(\sigma^2_i)=1$
and strong heteroscedasticity with $\Var(\sigma^2_i)=10$.
In this section we begin by summarizing the mild heteroscedastic
case.  The other cases are similar and we give some results for them
later.

To measure the loss in estimating $X$ due to using an estimate $\hat k$ 
instead of the optimal choice $k_{\text{ESA}}^*$ 
we use a relative estimation error (REE)  given by
\begin{align*}
	\text{REE}(\hat k) = \frac{\Vert\hat X(\hat k) - X \Vert_F^2}
{\Vert\hat X(k_{\text{ESA}}^*) - X \Vert_F^2} - 1.
\end{align*}
REE is zero if $\hat k$ is the best possible rank for the specific
data matrix shown, that is, if $\hat k$ is the same rank an oracle
would choose.

Let $m = \min(n, N)$ and the singular values of the data matrix 
$Y$ be $\sqrt n\lambda_1, \sqrt n \lambda_2, \cdots, \sqrt n \lambda_m$
in nonincreasing order. 
We compare BCV with $5$ representative methods reviewed in 
Section \ref{sec:prior}. They are: 
\begin{enumerate}
  \item PA: the permutation version of the parallel analysis \cite{buja1992remarks}.
The Gaussian version of \cite{horn1965rationale} has nearly identical
performance in our test cases.
  \item ED: the eigenvalue difference method \cite{onatski2010} which estimates the number 
    of factors as 
    $$\hat k = \max\{i \leq k_{\mathrm{max}}: \lambda_i^2 - \lambda_{i + 1}^2 \geq \delta\}$$
    where asymptotically $k_{\mathrm{max}}$ should be a slowly increasing function of $n$,
     and $\delta$ is calculated via a calibration method described in \cite{onatski2010}. 
If $\{i \leq k_{\mathrm{max}}: \lambda_i^2 - \lambda_{i + 1}^2 \leq \delta\}$ is empty
then we take $\hat k = 0$.
   \item ER: the eigenvalue ratio method \cite{ahn2013} which is the maximizer of 
     sequential eigenvalue ratios
     $$\hat k = \argmax_{0 \leq i \leq k_{\mathrm{max}}}\frac{\lambda_i^2}{\lambda_{i + 1}^2}$$
     where $\lambda_0^2 = \sum_{i = 1}^{m} \lambda_i^2/\log(m)$. 
     Also, $k_{\mathrm{max}}$ is suggested to be determined as 
     $\min(|\{j\geq 1: \lambda_j^2 \geq \sum_{i = 1}^m \lambda_i^2/m\}|, 0.1m)$.
   \item IC1: one of the rules based on information criteria developed in 
     \cite{bai2002determining}. It is the $k$ value that minimizes the criterion function
     $$\mathrm{IC}_1(k) = \log(V(k)) + k\Big(\frac{n + N}{nN}\Big)\log\Big(\frac{nN}{n + N}\Big)$$
     where $V(k) = \|Y - \hat Y(k)\|_F^2/(nN)$.
   \item NE: Nadakuditi and Edelman's information criteria based estimator 
     \cite{nadakuditi2008sample} which aims to estimate the number of weak factors in the white 
     noise model. Set
     $$t_i = N\Big[(N - i) \frac{\sum_{j = i + 1}^ N \lambda_j^4}
             {(\sum_{j = i + 1}^N \lambda_j^2)^2} - \Big(1 + \frac{N}{n}\Big)\Big]
             - \frac{N}{n},$$
and then choose
      $$\hat k = \argmin_{0 \leq i <\min(N, n)} 
      \Big[\frac{1}{2}\Big( \frac{n}{N}\Big)^2 t_i^2 + 2(i + 1)\Big].$$ 
  \end{enumerate}

Of these methods, ER and IC1 are designed for models with strong factors only.
ED does not require strong factors to work. NE has theoretical guarantees 
  for estimating the number of detectable weak factors in the white noise model. Finally, PA was 
  designed and tested under the small $N$ and large $n$ scenarios. 
  We want to compare the finite sized dataset performance of these methods in settings
  with both strong and weak factors.  In applications one cannot be sure that only
  the desired factor strengths are present. In an earlier version of the paper 
  \cite{owen2015}, we 
  also compared with Kaiser's rule \cite{kaiser1960application}, 
  which estimates the number of factors as 
  the number of eigenvalues of sample correlation matrix above $1$. However, Kaiser's rule 
  is likely to over estimate the number of factors and does not perform well. 
We also include in the
comparison the use of the true number of factors as well as
the oracle's number of factors 
$k_{\text{ESA}}^*$ defined in (\ref{optk}). 
Methods that choose a value closer to $k_{\text{ESA}}^*$, 
should attain a small error using ESA.

Figure \ref{methods_survival} shows for different methods, the proportion of simulations
with REE above certain values 
for the mild heteroscedastic case $\Var(\sigma_i^2) = 1$.
Figure \ref{fig:ESA-all} shows that BCV is overall best at recovering the
signal matrix $X$.
BCV is based on Perry's asymptotic Theorem~\ref{bcv_thm}. Figure~\ref{fig:ESA-large}
shows that BCV becomes far better than alternatives when we just compare
the larger sample sizes from each aspect ratio.
Figure~\ref{fig:ESA-small} shows that at smaller sample sizes NE is competitive with
BCV.  The large data case is more important given the recent emphasis on
large data problems.

Our goal is to find the best $k$ for ESA, but
the methods ED, ER. IC1 and NE are designed assuming that the SVD will be used to 
estimate the factors. To study them in the setting they were designed for,
we include Figure \ref{fig:SVD-all}, which 
calculates REE using SVD to estimate $\hat X(k)$ and compares with the oracle rank of 
SVD. For Figure \ref{fig:SVD-all}, the BCV method also uses
the SVD instead of ESA.  
Though the results in Table \ref{summary_table} (Appendix) suggest that SVD is in general not 
recommended for heteroscedastic noise data, if one does use SVD, BCV is still the 
best method for choosing $k$ to recover $X$.

The proportion of simulations with $\mathrm{REE} = 0$ (matching the oracle's rank)
for BCV was $51.6\%$, $75.1\%$, $28.1\%$ and $47.0\%$
in the four scenarios in Figure \ref{methods_survival}. BCV's percentage was always highest
among the six methods we used. The fraction of $\mathrm{REE}=0$ sharply increases
with sample size and is somewhat better for ESA than for SVD.

\newcommand{\panelsize}{0.42} 
\begin{figure}[t!]
\centering  
  \begin{subfigure}[b]{\panelsize\textwidth}
  \includegraphics[width = \textwidth]{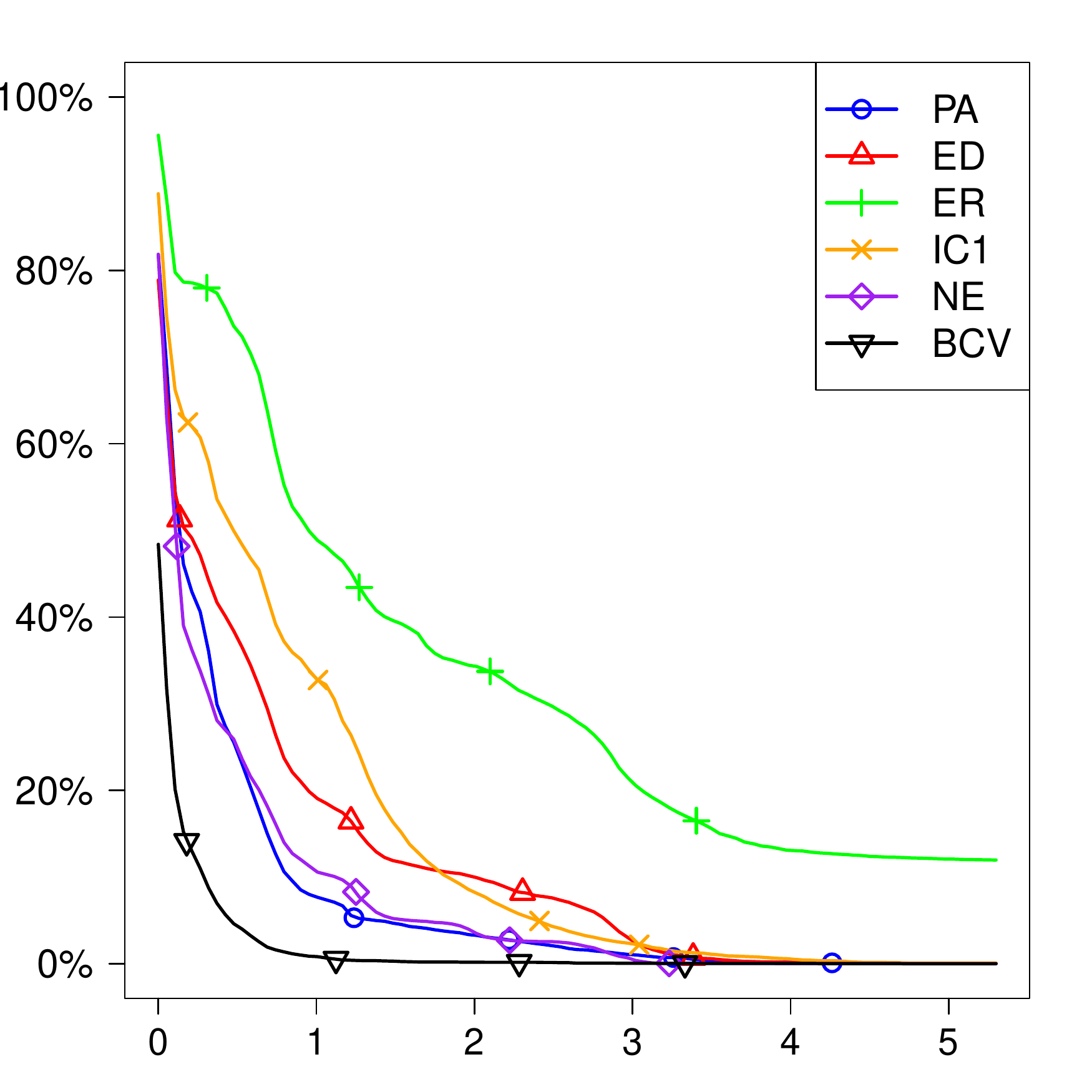}
  \caption{All datasets, ESA}
  \label{fig:ESA-all}
  \end{subfigure}
  \begin{subfigure}[b]{\panelsize\textwidth}
  \includegraphics[width = \textwidth]{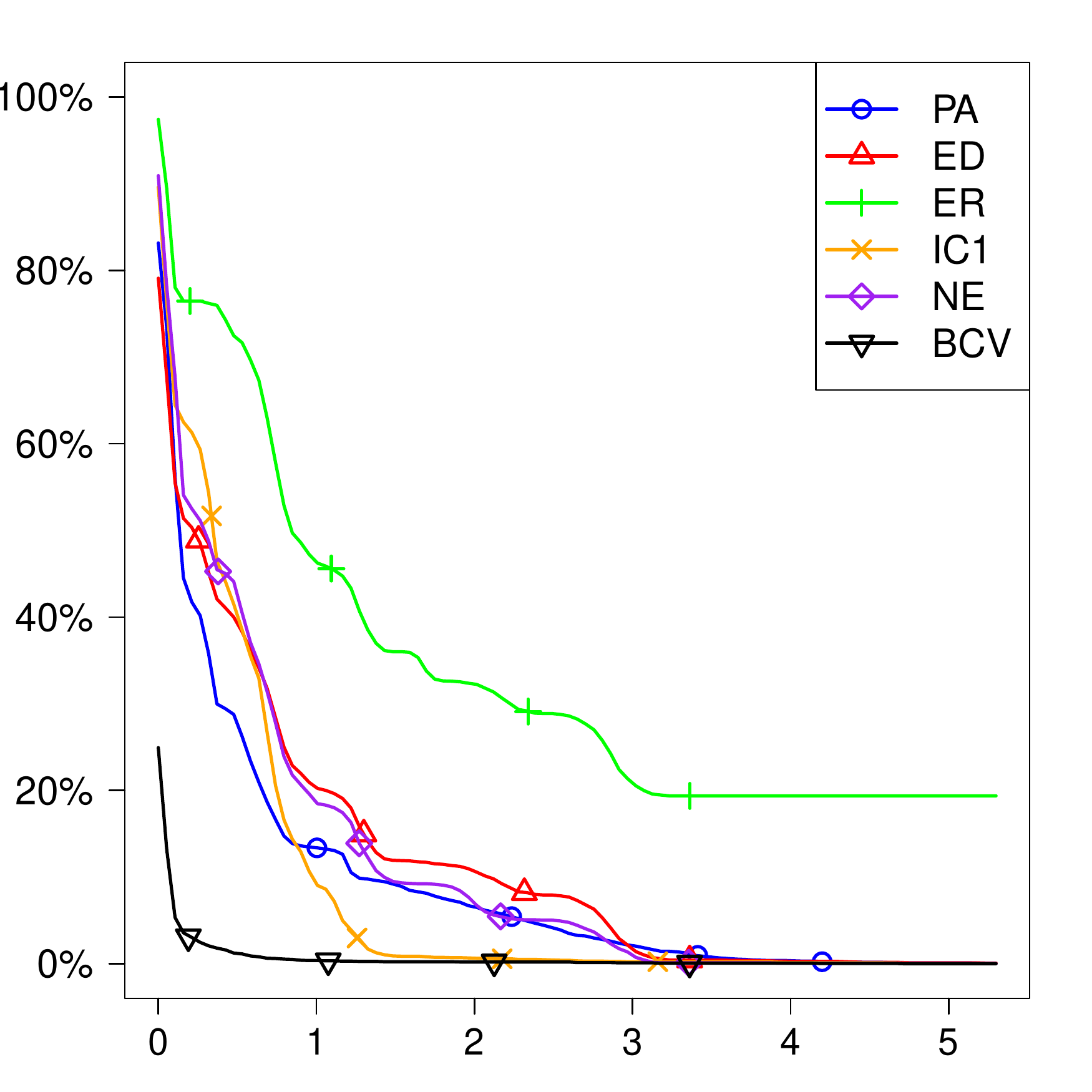}
  \caption{Large datasets only, ESA}
  \label{fig:ESA-large}
  \end{subfigure}
  \begin{subfigure}[b]{\panelsize\textwidth}
  \includegraphics[width = \textwidth]{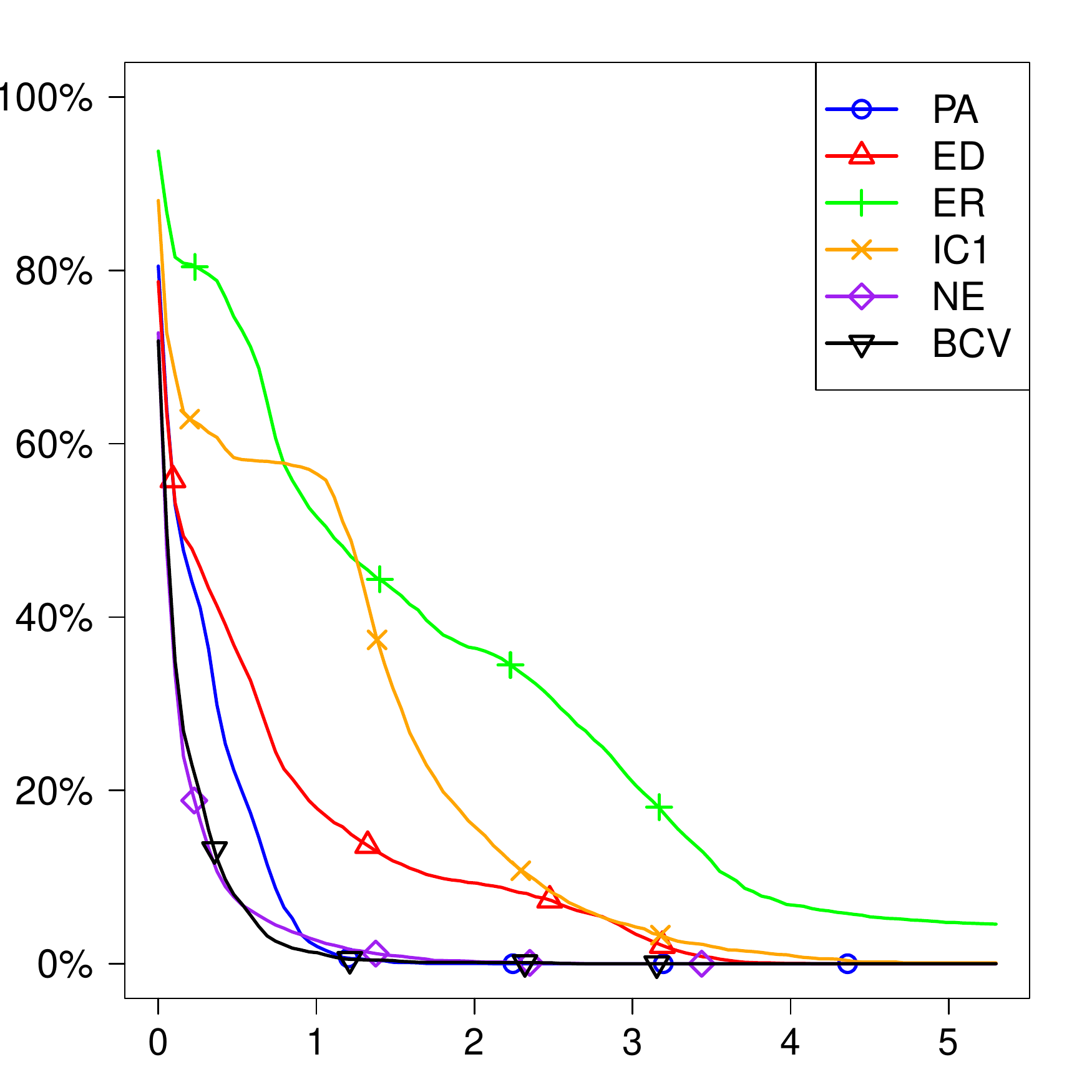}
  \caption{Small datasets only, ESA}
  \label{fig:ESA-small}
  \end{subfigure}
  \begin{subfigure}[b]{\panelsize\textwidth}
  \includegraphics[width = \textwidth]{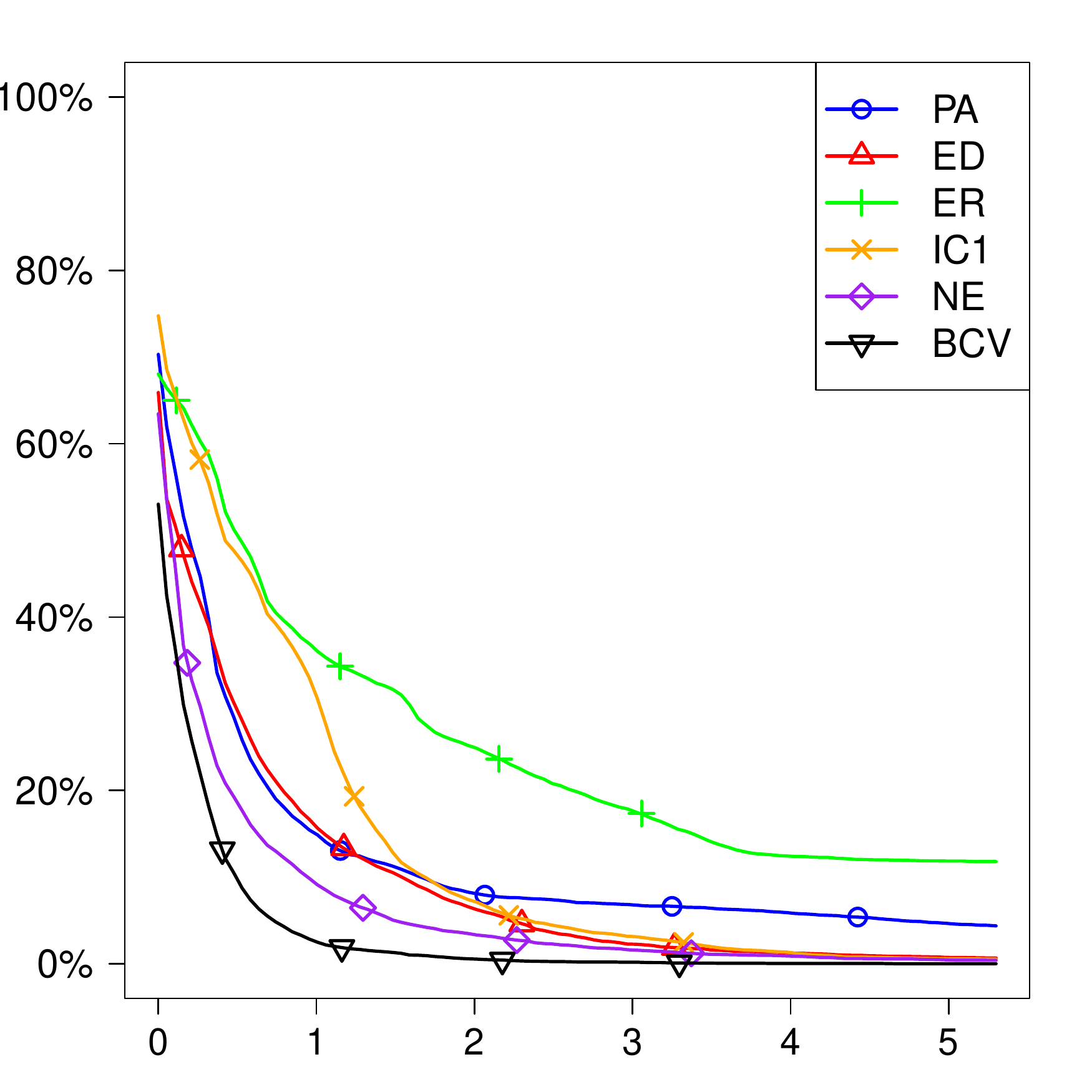}
  \caption{All datasets, SVD}
  \label{fig:SVD-all}
  \end{subfigure}
\caption{ REE survival plots:
 the proportion of samples with REE exceeding the number
on the horizontal axis. 
Figure \ref{fig:ESA-all}-\ref{fig:ESA-small}
are for REE calculating using the method ESA.
Figure \ref{fig:ESA-all} shows all $6000$ samples.
Figure \ref{fig:ESA-large} shows only the $3000$ simulations of larger matrices 
of each aspect ratio.
Figure \ref{fig:ESA-small} shows only the $3000$ simulations of smaller matrices. 
For comparison, 
Figure \ref{fig:SVD-all} is the REE plot for all samples calculating REE 
using the method SVD.
}
\label{methods_survival}
\end{figure}

Table~\ref{tab:overall} briefly summarizes the REE values for all 
three noise variance cases. It shows the worst case REE over all the 
$10$ matrix sizes and $6$ factor strength scenarios. 
As the variance of $\sigma^2_i$ rises it becomes more
difficult to attain a small REE.  
BCV has substantially smaller worst case REE for heterscedastic noise than all 
other methods, 
but is slightly worse than NE for the white noise case. This is not surprising as 
NE is designed for the white noise model.

\begin{table}[t]
\centering 
\begin{tabular}{@{}ccccccc@{}}
  \toprule 
  $\Var(\sigma^2_i)$ & PA & ED & ER & IC1 & NE & BCV \\  \midrule 
\phz0 & 1.99 & 1.41 & 49.61 & 1.13 & 0.12 & 0.29 \\ 
\phz1 & 2.89 & 2.42 & 25.02 & 3.11 & 2.45 & 0.37 \\ 
  10 & 3.66 & 2.28 & 15.62 & 4.46 & 2.10 & 0.62 \\ 
   \bottomrule 
\end{tabular}
\caption{\label{tab:overall}
Worst case REE values for each method of choosing $k$
for white noise and two heteroscedastic noise settings. 
}
\end{table}

\begin{figure}
\centering
\includegraphics[width=0.95\textwidth]{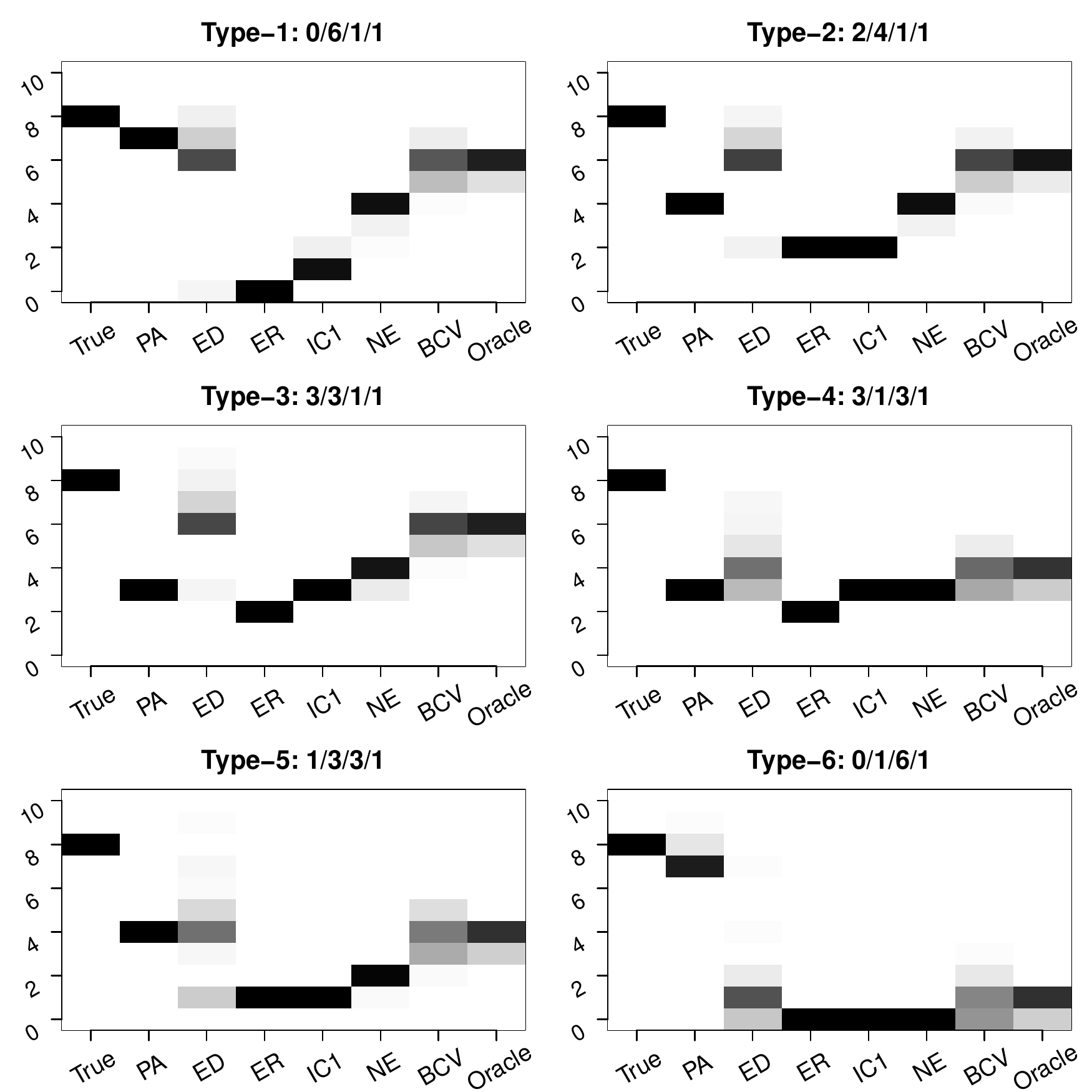}
\caption{The distribution of $\hat k$ for each factor strength case
  when the matrix size is $5000 \times 100$. 
  The y axis is $\hat k$. 
  Each image depicts $100$ 
simulations with counts plotted in grey scale (larger equals darker). 
 For different scenarios, the factor strengths 
  are listed as the number of ``strong/useful/harmful/undetectable'' factors in the title of 
each subplot.
The true $k$ is always $k_0 = 8$. The ``Oracle'' method corresponds to $k_\text{ESA}^*$.}
\label{rank_distribution}
\end{figure}

To better understand the differences among the methods, 
we compare them directly in estimating 
the number of factors with the oracle. 
As an example, 
Figure \ref{rank_distribution} plots 
the distribution of $\hat k$ for all methods and all $6$ cases,
on $5000\times 100$ data matrices with $\Var(\sigma_i^2) = 1$. 
The results of other cases are summarized in 
Tables \ref{compare_k_table} and \ref{compare_k_table_con} 
in the Appendix. 
In Figure \ref{rank_distribution}, 
BCV closely tracks the oracle. For other methods, ED performs the 
best in estimating the oracle rank, though it is more variable and less 
accurate than BCV. ER is the most conservative method, trying to estimate 
at most the number of strong factors. IC1 also tries to estimate the 
number of strong factors, but is less conservative than ER. NE estimates 
some number between the number of strong factors and the number of 
useful (including strong) factors. 
PA has trouble identifying the useful weak factors when strong factors
are present, and also has trouble rejecting the detectable but not
useful factors in the hard case with no strong factor. This is due 
the fact that PA is using the sample correlation matrix which has 
a fixed sum of eigenvalues, thus the magnitude of the each eigenvalue is
influenced by every other one.

Tables \ref{compare_k_table} and \ref{compare_k_table_con}
in the Appendix provide more details of the simulation results
for this mildly heteroscedastic case $\Var(\sigma_i^2) = 1$. 
We can see that some methods behave very differently for 
different sized datasets. For example, IC1 is very non-robust and 
sharply over-estimates the number of factors for small datasets, ED will tend to estimate 
only the number of strong factors when the aspect ratio $\gamma$ is small. 
Overall, BCV has the most robust and accurate performance in estimating 
$k_{\mathrm{ESA}}^*$ of the methods we investigated.

\section{Real Data Example}\label{sec:data}
We investigate a real data example to show how our method works in practice. The observed matrix $Y$ is $15 \times 8192$, where each row is a chemical element and each column represents a position on a $64 \times 128$ map of a meteorite. 
We thank Ray Browning for providing this data. 
Similar data are discussed in \cite{paque1990quantitative}.
Each entry in $Y$ is the amount of a chemical element at a grid point. The task is to analyze the distribution patterns of the chemical elements on that meteorite, helping us to further understand the composition.

A factor structure seems reasonable for the elements as various
compounds are distributed over the map.
The amounts of some elements such as Iron and 
Calcium are on a much larger scale than some other elements like Sodium and 
Potassium, and so it is necessary to assume a heteoroscedastic noise model as 
(\ref{model}). We center the data for each element before applying our method.

\begin{figure}[t]
	\center
\includegraphics[width=0.6\textwidth]{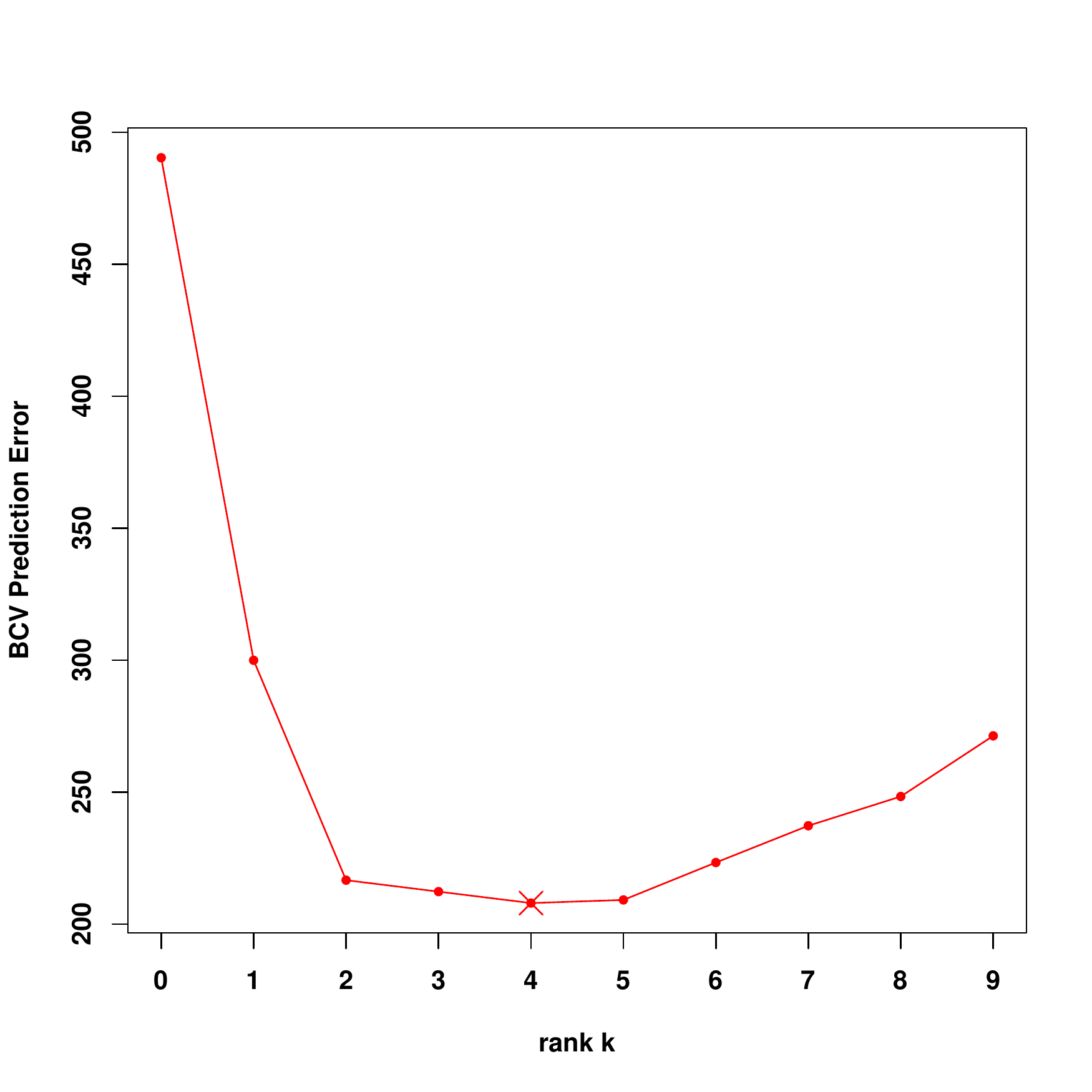}
\caption{BCV prediction error for the meteorite. The BCV partitions have been repeated
200 times. The solid red line is the average over all held-out blocks, with the cross 
marking the minimum BCV error.}
\label{bcv_meteo}

\end{figure}

\begin{figure}
	\center
\includegraphics[width=0.95\textwidth]{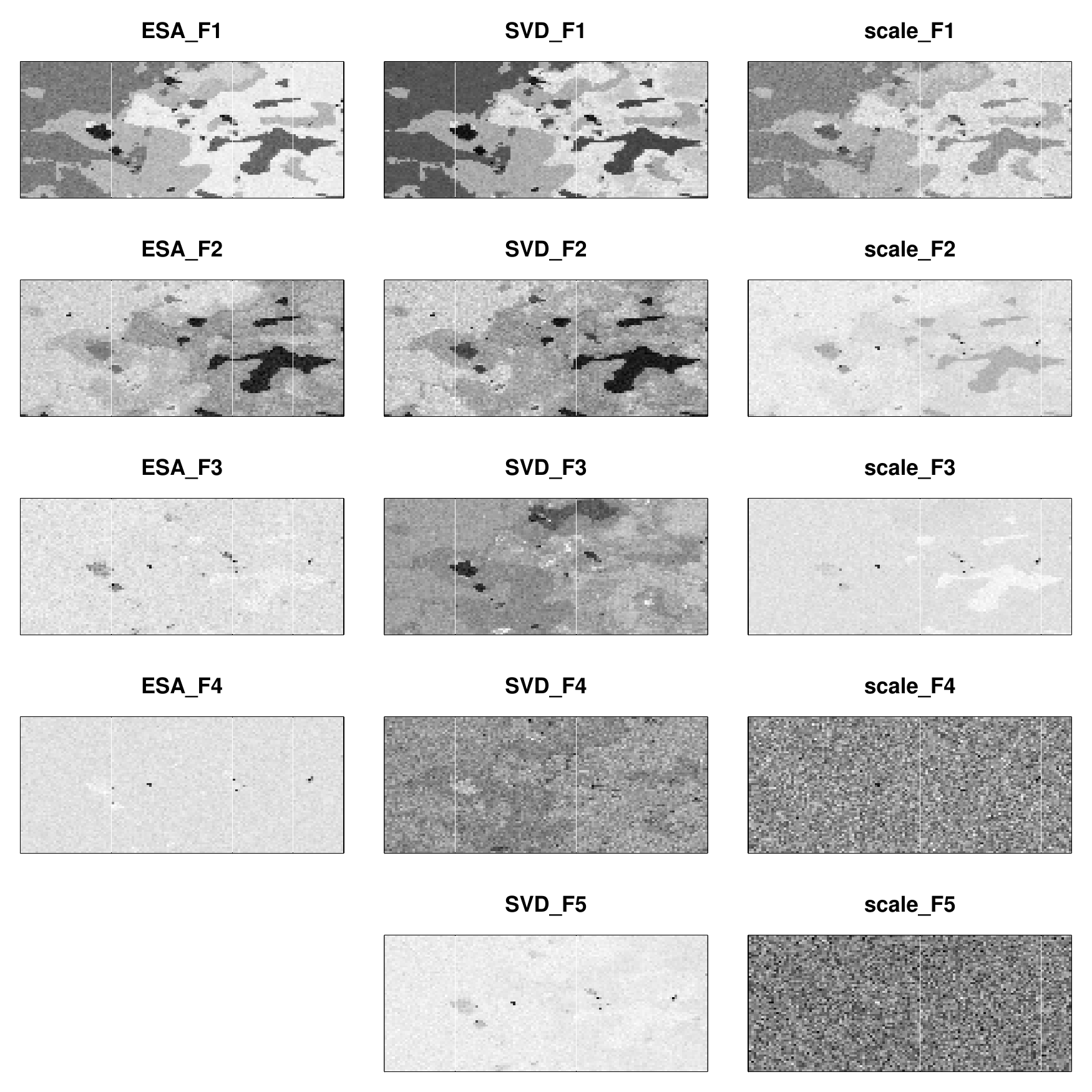}
\caption{Distribution patterns of the estimated factors. The first column has the four factors found by ESA. The second column has the top five factors found by applying SVD on the unscaled data. The third column has the top five factors found by applying SVD on scaled data in which each element has been standardized. The values are plotted in grey scale, and a darker color indicates a higher value.}
\label{meteo_pos}

\end{figure}

BCV choose $k=4$ factors, while PA chooses $k=3$.
Figure \ref{bcv_meteo} plots the BCV error for each rank, showing 
that among the selected factors, the first two factors are much more influential than 
the last two. The first column of Figure \ref{meteo_pos} plots the four factors ESA 
has found at their positions. They represents four clearly different patterns. 

As a comparison, we also apply a straight SVD on the centered data with and without 
standardization to analyze the hidden structure. The second and third columns of 
Figure \ref{meteo_pos} shows the first five factors of the locations that SVD 
finds for the original and scaled data respectively. If we do not scale
the data, then the
factor (F5) showing the concentration of Sulfur on some specific locations strangely 
comes after the factor (F4) which has no apparent pattern;
F5 would have been neglected in a model of three or four factors
as BCV or PA suggest. 

\begin{figure}
	\center
	\includegraphics[width = 0.95\textwidth]{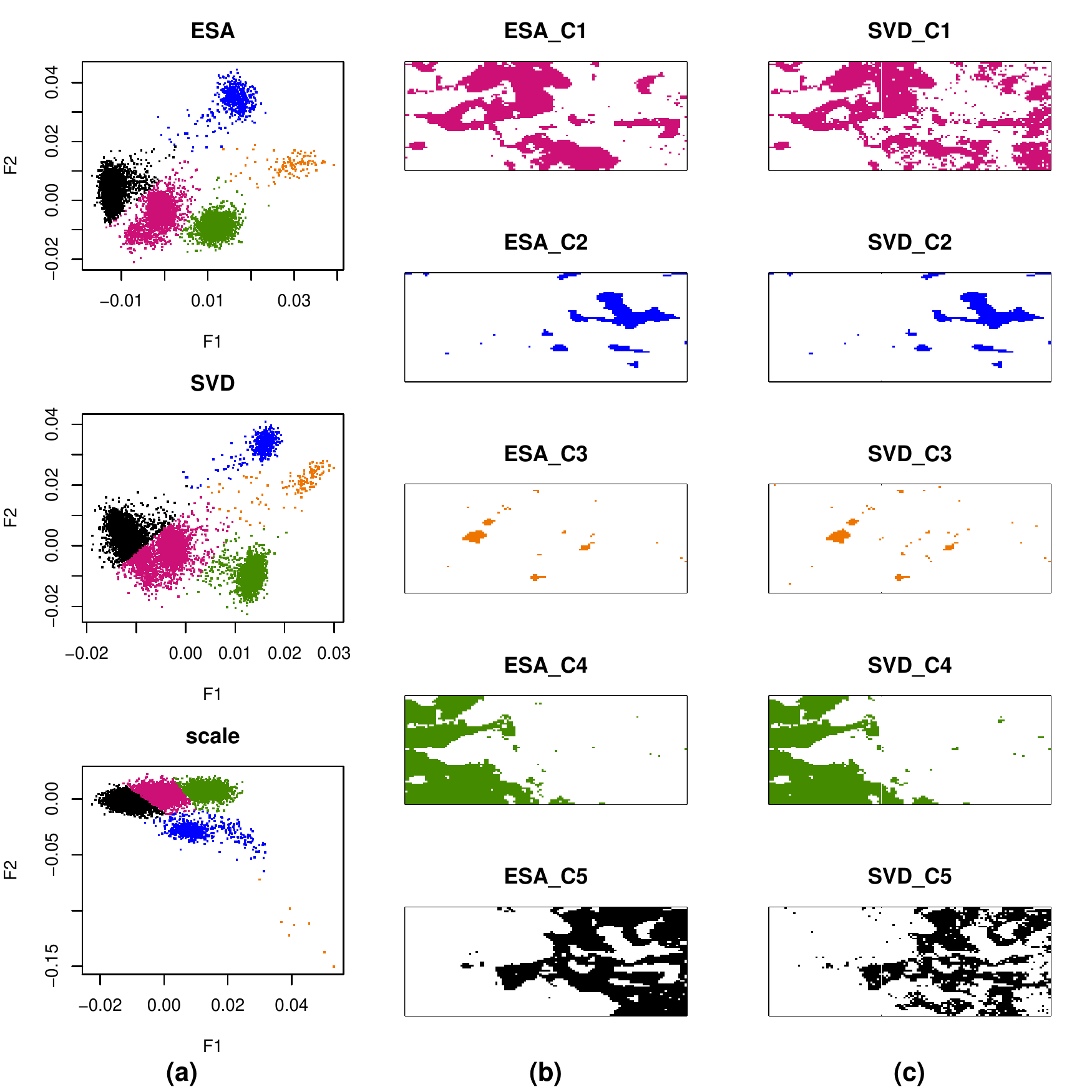}
	\caption{Plots of the first two factors and the location clusters. The three plots of column (a) are the scatter plots of pixels for the first two factors found by the three methods: ESA, SVD on the original data and SVD on normalized data. The coloring shows a k-means clustering result for $5$ clusters. Column (b) has the five clustered regions based on the first two factors of ESA. Column (c) has the five clustered regions based on the first two factors of SVD on the original data after centering. The same color represents the same cluster.}
\label{result_1}
\end{figure}

Paque et al.{}\cite{paque1990quantitative} 
investigate this sort of data by clustering the pixels
based on the values of the first two factors of a factor analysis.
We apply such a clustering in Figure~\ref{result_1}.
Column (a) shows the resulting clusters.
The factors found by ESA clearly divide the locations into five clusters, while 
the factors found by an SVD on the original data blur the boundary between
clusters 1 and 5.  An SVD on normalized data (third plot in column (a))
blurs together three of the clusters.
Columns (b) and (c) of Figure \ref{result_1} show the quality of 
clustering using k-means based on the first two plots of Column (a). Clusters, 
especially C1 and C5, have much clearer boundaries and are less noisy if we are 
using ESA factors than using SVD factors. 
A $k$-means clustering depends on the starting points.  For the ESA data the 
clustering was stable.  For SVD the smallest group C3 was sometimes merged
into one of the other clusters; we chose a clustering for SVD that preserved C3.

In this data the ESA based factor analysis found factors
that, visually at least, seem better. They have better spatial
coherence, and they provide better clusters than the SVD
approaches do. For data of this type it would be reasonable to use 
spatial coherence of the latent variables to improve the fitted model. 
Here we have used spatial coherence as an informal confirmation that 
BCV is making a reasonable choice, which we could not
do if we had exploited spatial coherence in estimating our factors.


\subsection{AGEMAP data}

The meteorite data is the second of two real world data sets
that we have tried BCV on.
The first was the AGEMAP data used to study the
LEAPP algorithm \cite{sun2012}.
There, instead of a gold standard of a known signal matrix,
the notion of ground truth is supplied by the idea that
a better estimate of the signal in expression matrices for $16$ different tissues
should lead to greater overlap among the genes declared significant in those tissues.
This is an indirect gold standard like the idea of positive controls in 
\cite{gagnon2012using}.
The LEAPP algorithm used parallel analysis as implemented
in the SVA package of \cite{leek2008general}.


Placing BCV in LEAPP for the AGEMAP data
yields a result similar to PA on the correlation matrix
but is somewhat less effective than PA with the covariance matrix.
All three are fairly close and all three gave better overlap than SVA did.

We do not understand why BCV failed to improve the overlap
measure for the AGEMAP data. Here are some possibilities:
We simulated Gaussian data using guidance from mostly
Gaussian RMT, and the real data might not have been close enough
to Gaussian. The noise covariance in AGEMAP might not have
been nearly diagonal.
There may not have been enough harmful factors in the AGEMAP data
for the differences to be observed.  LEAPP may be robust to missing
weak factors. Finally, there is no reason to expect
that one method will be closer to an oracle on every data set.

\section{Conclusion}\label{sec:conclusion}
In this paper, we have developed a bi-cross-validation algorithm to choose 
the number of factors in a heteroscedastic factor analysis and an early
stopping alternation to estimate the model.
Guided by random matrix theory, we have constructed a battery of test
scenarios and found that stopping at three iterations is very effective.
Using that early stopping rule we find that
our bi-cross-validation proposal produces better recovery
of the underlying signal matrix
than other widely used methods. 
It also improves markedly with sample size.

\section*{Acknowledgments}

This work was supported by the National Science Foundation
under grants DMS-1407397 and DMS-0906056. We thank the
reviewers for comments that lead to an improved paper.

\bibliographystyle{plain}
\bibliography{ref.bib}

\section*{Appendix}

\subsection*{A.1: Simulation test cases}

Our model is a low rank signal plus heteroscedastic
noise.  The formulation $Y=\sqrt{n}UDV^\tran+E$
does not make it easy to take account of random matrix theory.
We write our model as
\begin{align}\label{color}
	Y = \Sigma^{1/2}(\sqrt{n}UDV^\tran  + E)
\end{align}
where $\Sigma = \diag (\sigma_1^2, \sigma_2^2, \cdots, \sigma_N^2)$ and $\sqrt{n}UDV^\tran $ is the SVD for $\Sigma^{-1/2}X$.
For constant $\sigma_i^2=\sigma^2$ RMT can be used
to choose the entries of
$D = \diag (d_1, d_2, \cdots, d_{k_0})$ 
where $d_1 > d_2 > \cdots > d_{k_0} > 0$.

A straightforward implementation of~\eqref{color} would
have uniformly distributed $U$. In that case however the
mean square signal per row would be simply proportional
to the noise mean square per row.  We think this would
make the problem unrealistically easy: the relative sizes of
the noise variances would be well estimated by corresponding
sample variances within rows.  Our simulation chooses 
a non-uniform $U$ in order to decouple the mean square
signal of the rows from the mean square noise in the rows.
Below are the rules for generating the simulated data. 

\subsubsection*{Generating the noise}
Recall that the noise matrix is $\Sigma^{1/2} E$. The steps are as follows.
\begin{enumerate}
	\item $E= \left(e_{ij}\right)_{N \times n}$: here $e_{ij} \simiid \dnorm(0, 1)$. 
	\item $\Sigma=\diag(\sigma_1^2,\dots,\sigma_N^2)$: 
$\sigma_i^2\simiid\mathrm{InvGamma}(\alpha, \beta)$. 
Therefore $\E(\sigma_i^2) = {\beta}/{\alpha - 1}$ and 
$\Var(\sigma_i^2) = {\beta^2}/{(\alpha - 1)^2(\alpha - 2)}$. 
Parameters $\alpha$ and $\beta$ are chosen so that
$\E(\sigma_i^2) = 1$.  We consider two heteroscedastic noise cases: 
$\Var(\sigma_i^2) = 1$ and $\Var(\sigma_i^2) = 10$.
We also include a homoscedastic case with all $\sigma^2_i=1$. 
\end{enumerate}

\subsubsection*{Generating the signal}
The signal matrix is $X=\sqrt{n} \Sigma^{1/2} UDV^\tran$, where $\Sigma$ is the
same matrix used to generate the noise.
Entries in $D$ specify the strength of signals of the reweighted matrix $\Sigma^{-1/2}X$. As we discussed in 
Section \ref{sec:testcases},
for high-dimensional white noise models \cite{2009arXiv0909.3052P}, there are two thresholds of signal strength for truncated SVD: a detection threshold and an estimation threshold. 
From 
\cite{2009arXiv0909.3052P} 
the detection threshold is 
$\mu_F  = \sqrt\gamma$ and 
the estimation threshold is 
$$\mu_F^*  = \frac{1 + \gamma}{2} + \sqrt{\left(\frac{1 + \gamma}{2}\right)^2 + 3\gamma},$$
in the homoscedastic $\sigma=1$ case.  
Recall that based on the asymptotic thresholds, our four categories for a 
dataset are roughly:
\begin{compactenum}[\quad a)]
\item Undetectable, $d_i^2<\mu_F$,
\item Harmful, $\mu_F<d_i^2<\mu_F^*$,
\item Useful, $\mu_F^*<d_i^2=o(1)$, and 
\item Strong, $d_i^2 \sim O(N)$.
\end{compactenum}

\medskip
The signal simulation is as follows.
\begin{enumerate}
\item
We include the $6$ scenarios
from Table \ref{factor_strength_table}. For the $d_i^2$ values we take, 
the strong factors takes values at $1.5N$, $2.5N$, $3.5N$, $\cdots$. The 
useful factors takes values at $1.5\mu_F^\star$, $2.5\mu_F^\star$, 
$3.5\mu_F^\star$, $\cdots$. The harmful factors takes values at equally spaced interior
points of the interval $[\mu_F, \mu_F^\star]$ and the undetectable 
factors takes values at equally spaced interior points of the interval $[0, \mu_F]$.
	\item $U$ and $V$: First $V$ is sampled uniformly from the Stiefel manifold  $V_k(\R^n)$. See Appendix $A.1.1$ in \cite{2009arXiv0909.3052P} for a suitable algorithm. Then an intermediate matrix $U^*$ is sampled uniformly from 
the Stiefel manifold $V_k(\R^N)$. 
Using the previously generated $V$ and $\Sigma$ we solve
$$\Sigma^{-1/2}U^*DV^\tran  = U\tilde D\tilde V^\tran $$
for $U$. Now $U$ is nonuniformly distributed on on the Stiefel manifold
in such a way that rows of $U$ with large $L^2$ norm are not 
necesarily those with large $\sigma^2_i$. 
\end{enumerate}

\subsubsection*{Data dimensions}
We consider $5$ different $N/n$ ratios: $0.02, 0.2, 1, 5, 50$ and for each ratio consider a small matrix size and a larger matrix size, thus there are in total $10$ $(N, n)$ pairs. 
The specific sample sizes appear 
at the top of Table~\ref{ESA_table}.
In total there are $6\times 3\times 5\times 2=180$ scenarios.
Each was simulated $100$ times, for a total of $18{,}000$
simulated data sets.

\subsection*{A.2: Early stopping}

To study the effects of early stopping, we investigated the
cases from Appendix A.1, varying the number $k$ of factors
and varying the number $m$ of steps.
In these simulations we know the true signal $X$ and
so we can measure the errors.
We use the six measurements below to study the effectiveness
of ESA with $m=3$:
\begin{enumerate}
\item $\text{Err}_X(m = 3)\big/\text{Err}_X(m = m_{\text{Opt}})$:
		
this compares $m=3$ to the optimal $m$ defined in (\ref{optm}). 
\item $\text{Err}_X(m = 3)\big/\text{Err}_X(m=1)$:
		
this measures the advantage of 
ESA beyond PCA.

\item $\text{Err}_X(m = 3)\big/\text{Err}_X(m=50)$:
		
this measures the advantage of stopping early, using $m=50$ as proxy
for iteration to convergence.

\item $\text{Err}_X(m = 3)\big/\text{Err}_X(\text{SVD})$:

this compares ESA to the truncated SVD one would
do for homoscedastic data.

\item $\text{Err}_X(m = 3)\big/\text{Err}_X(\text{QMLE})$:

 this compares ESA to the quasi maximum likelihood method, which is solved using the EM 
 algorithm with principal component estimates as starting values.

\item  $\text{Err}_X(m = 3)\Big/\text{Err}_X(\text{oSVD})$:

this compares ESA to the truncated SVD that an oracle which knew $\Sigma$
could use on $\Sigma^{-1/2}Y$.
It measures the relative inaccuracy in $\hat X$ arising from the 
inaccuracy of $\hat\Sigma$. 

\end{enumerate}
For QMLE, $R$ and $\Sigma$ are estimated via maximizing the quasi-loglikelihood 
\cite{bai2012}:
\begin{align}\label{eq:qmle}
  - \frac{1}{2N} \log \mathrm{det}\left(RR^T + \Sigma\right) - \frac{1}{2N}
  \mathrm{tr}\left\{\frac{YY^T}{n} \left[RR^T + \Sigma\right]^{-1}\right\}
\end{align}
then $\hat L$ is estimated via a generalized linear regression of $Y$ on $\hat R$ with 
estimated variance $\hat \Sigma$ and $\hat X = \hat R \hat L$.

\begin{table}[t]
\centering 
\scalebox{0.9}{
\begin{tabular}{@{}lrrr@{}}
  \toprule 
  \multirow{2}{3cm}{Measurements}  & White Noise & \multicolumn{2}{c}{Heteroscedastic Noise} \\  
 \cmidrule(lr){2-2} 
 \cmidrule(lr){3-4} 
 & $\mathrm{Var}(\sigma_i^2) = 0$ & $\mathrm{Var}(\sigma_i^2) =1$ & $\mathrm{Var}(\sigma_i^2) = 10$ \\ 
 \midrule 
$\frac{\text{Err}_X(m = 3)}{\text{Err}_X(m = m_{\text{Opt}})}$ & $1.01\pm0.01$ & $1.00\pm0.01$ & $1.00\pm0.01$ \\ 
   \addlinespace[0.25cm] 
$\frac{\text{Err}_X(m = 3)}{\text{Err}_X(m=1)}$ & $0.93\pm0.09$ & $0.90\pm0.11$ & $0.89\pm0.12$ \\ 
   \addlinespace[0.25cm] 
$\frac{\text{Err}_X(m = 3)}{\text{Err}_X(m=50)}$ & $0.87\pm0.21$ & $0.87\pm0.21$ & $0.87\pm0.21$ \\ 
   \addlinespace[0.25cm] 
$\frac{\text{Err}_X(m = 3)}{\text{Err}_X(\text{SVD})}$ & $1.03\pm0.06$ & $0.81\pm0.20$ & $0.75\pm0.22$ \\ 
   \addlinespace[0.25cm] 
$\frac{\text{Err}_X(m = 3)}{\text{Err}_X(\text{QMLE})}$ & $1.02\pm0.05$ & $0.95\pm0.15$ & $0.91\pm0.19$ \\ 
   \addlinespace[0.25cm] 
$\frac{\text{Err}_X(m = 3)}{\text{Err}_X(\text{oSVD})}$ & $1.03\pm0.06$ & $1.03\pm0.07$ & $1.03\pm0.08$ \\ 
   \bottomrule 
\end{tabular}
}
\caption{ESA using six measurements. For each of Var$(\sigma_i^2) = 0, 1$ and $10$, the average for every measurement is the average over $10 \times 6 \times 100 = 6000$ simulations, and the standard deviation is the standard deviation of these 6000 simulations.} 
\label{summary_table}
\end{table}

Table \ref{summary_table} summarizes the mean and standard deviation of each 
measurement over $6000$ simulations each, for $\Var(\sigma_i^2) = 0$, $1$ and 
$10$. 
Row 1 shows that ESA stopping at $m=3$ steps was almost identical to 
stopping at the unknown optimal $m$ in terms of the oracle estimating 
error, as the mean is nearly $1$ and the standard deviation is negligible. 
Row 2 indicates that taking $m = 3$ steps brought
an improvement compared with PCA  
(SVD on standardized data).
Row 3 shows that taking $m=3$ brought an improvement compared
to using $m=50$, our proxy for iterating to convergence to the local minimum of loss. 
The latter is
highly variable.
Row 4 shows that truncated SVD is better than ESA 
when the noise is homoscedastic. But even a noise level as small
as $\Var(\sigma^2_i)=\E(\sigma^2_i)=1$ reverses the preference sharply. 
Row 5 shows that ESA beats QMLE on average for the heteroscedastic noise case, 
though the latter has theoretical guarantee for the strong factor scenario.
Row 6 shows that an oracle which knew $\Sigma$ and used it to
reduce the data to the homoscedastic case would gain only $3$\%
over ESA. 


\begin{table}[ht]
\centering
\scalebox{0.55}{
\begin{tabular}{@{}crrrrrrrrrr@{}}
  \toprule
    \multirow{2}{2cm}{Factor Scenario}  & \multicolumn{2}{c}{$\gamma = 0.02$} & \multicolumn{2}{c}{$\gamma = 0.2$} &\multicolumn{2}{c}{$\gamma = 1$} & \multicolumn{2}{c}{$\gamma = 5$} & \multicolumn{2}{c}{$\gamma = 50$} \\ 
  \cmidrule(lr){2-3} 
 \cmidrule(lr){4-5} 
 \cmidrule(lr){6-7} 
 \cmidrule(lr){8-9} 
 \cmidrule(lr){10-11} 
 & (20, 1000) & (100, 5000) & (20, 100) & (200, 1000) & (50, 50) & (500, 500) & (100, 20) & (1000, 200) & (1000, 20) & (5000, 100) \\ 
 \addlinespace[0.3cm] 
    \midrule 
 \addlinespace[0.3cm] 
 \multicolumn{3}{l}{$\text{Err}_X(m = 3)\Big/\text{Err}_X(m = m_{\text{Opt}})$}\\ 
 \addlinespace[0.2cm] 
Type-1 & 1.011 & 1.000 & 1.011 & 1.000 & 1.004 & 1.000 & 1.003 & 1.000 & 1.000 & 1.000 \\ 
  Type-2 & 1.013 & 1.002 & 1.012 & 1.001 & 1.006 & 1.000 & 1.004 & 1.000 & 1.000 & 1.000 \\ 
  Type-3 & 1.016 & 1.006 & 1.014 & 1.005 & 1.010 & 1.000 & 1.002 & 1.000 & 1.000 & 1.000 \\ 
  Type-4 & 1.002 & 1.002 & 1.009 & 1.001 & 1.008 & 1.000 & 1.006 & 1.000 & 1.000 & 1.000 \\ 
  Type-5 & 1.008 & 1.001 & 1.011 & 1.001 & 1.007 & 1.000 & 1.006 & 1.000 & 1.000 & 1.000 \\ 
  Type-6 & 1.007 & 1.000 & 1.011 & 1.000 & 1.006 & 1.000 & 1.003 & 1.000 & 1.001 & 1.000 \\ 
      \midrule 
 \addlinespace[0.3cm] 
 \multicolumn{3}{l}{$\text{Err}_X(m = 3)\Big/\text{Err}_X(m=1)$}\\ 
 \addlinespace[0.2cm] 
Type-1 & 0.900 & 0.936 & 0.913 & 0.957 & 0.924 & 0.977 & 0.967 & 0.987 & 0.995 & 0.998 \\ 
  Type-2 & 0.819 & 0.626 & 0.844 & 0.680 & 0.833 & 0.785 & 0.942 & 0.909 & 0.990 & 0.987 \\ 
  Type-3 & 0.827 & 0.613 & 0.840 & 0.616 & 0.801 & 0.739 & 0.925 & 0.887 & 0.987 & 0.984 \\ 
  Type-4 & 0.781 & 0.723 & 0.837 & 0.751 & 0.864 & 0.833 & 0.947 & 0.926 & 0.990 & 0.990 \\ 
  Type-5 & 0.854 & 0.789 & 0.904 & 0.834 & 0.911 & 0.899 & 0.962 & 0.956 & 0.993 & 0.994 \\ 
  Type-6 & 0.987 & 0.993 & 0.997 & 0.996 & 0.997 & 0.998 & 0.999 & 0.999 & 0.999 & 1.000 \\ 
      \midrule 
 \addlinespace[0.3cm] 
 \multicolumn{3}{l}{$\text{Err}_X(m = 3)\Big/\text{Err}_X(m=50)$}\\ 
 \addlinespace[0.2cm] 
Type-1 & 0.441 & 0.802 & 0.473 & 0.985 & 0.759 & 1.000 & 0.590 & 1.000 & 1.000 & 1.000 \\ 
  Type-2 & 0.472 & 0.839 & 0.486 & 0.984 & 0.765 & 1.000 & 0.605 & 1.000 & 1.000 & 1.000 \\ 
  Type-3 & 0.501 & 0.918 & 0.463 & 0.994 & 0.751 & 1.000 & 0.626 & 1.000 & 1.000 & 1.000 \\ 
  Type-4 & 0.560 & 0.975 & 0.541 & 0.989 & 0.899 & 1.000 & 0.854 & 1.000 & 1.000 & 1.000 \\ 
  Type-5 & 0.604 & 0.907 & 0.671 & 0.992 & 0.821 & 1.000 & 0.842 & 1.000 & 1.000 & 1.000 \\ 
  Type-6 & 0.947 & 0.982 & 0.981 & 0.999 & 0.988 & 1.000 & 0.997 & 1.000 & 1.000 & 1.000 \\ 
      \midrule 
 \addlinespace[0.3cm] 
 \multicolumn{3}{l}{$\text{Err}_X(m = 3)\Big/\text{Err}_X(\text{SVD})$}\\ 
 \addlinespace[0.2cm] 
Type-1 & 0.638 & 0.348 & 0.740 & 0.366 & 0.722 & 0.466 & 0.882 & 0.727 & 0.977 & 0.966 \\ 
  Type-2 & 0.785 & 0.450 & 0.829 & 0.451 & 0.749 & 0.525 & 0.898 & 0.754 & 0.980 & 0.972 \\ 
  Type-3 & 0.870 & 0.611 & 0.896 & 0.548 & 0.772 & 0.599 & 0.903 & 0.791 & 0.983 & 0.976 \\ 
  Type-4 & 0.872 & 0.810 & 0.923 & 0.809 & 0.893 & 0.872 & 0.960 & 0.942 & 0.991 & 0.990 \\ 
  Type-5 & 0.704 & 0.542 & 0.798 & 0.552 & 0.770 & 0.605 & 0.888 & 0.779 & 0.978 & 0.972 \\ 
  Type-6 & 0.935 & 0.906 & 0.972 & 0.925 & 0.971 & 0.943 & 0.985 & 0.966 & 0.993 & 0.991 \\ 
      \midrule 
 \addlinespace[0.3cm] 
 \multicolumn{3}{l}{$\text{Err}_X(m = 3)\Big/\text{Err}_X(\text{QMLE})$}\\ 
 \addlinespace[0.2cm] 
Type-1 & 0.915 & 0.633 & 0.966 & 0.677 & 0.985 & 0.858 & 0.997 & 0.988 & 1.000 & 1.000 \\ 
  Type-2 & 1.104 & 0.672 & 1.058 & 0.725 & 1.000 & 0.863 & 0.999 & 0.989 & 1.000 & 1.000 \\ 
  Type-3 & 1.199 & 0.826 & 1.129 & 0.766 & 1.008 & 0.878 & 0.997 & 0.990 & 1.000 & 1.000 \\ 
  Type-4 & 1.035 & 0.991 & 1.033 & 0.954 & 1.005 & 0.973 & 1.002 & 0.997 & 1.000 & 1.000 \\ 
  Type-5 & 0.966 & 0.661 & 0.996 & 0.744 & 0.989 & 0.885 & 0.998 & 0.991 & 1.000 & 1.000 \\ 
  Type-6 & 0.971 & 0.912 & 0.993 & 0.942 & 0.999 & 0.974 & 0.999 & 0.999 & 1.000 & 1.000 \\ 
      \midrule 
 \addlinespace[0.3cm] 
 \multicolumn{3}{l}{$\text{Err}_X(m = 3)\Big/\text{Err}_X(\text{oSVD})$}\\ 
 \addlinespace[0.2cm] 
Type-1 & 1.029 & 0.994 & 1.064 & 0.998 & 1.036 & 1.001 & 1.026 & 1.001 & 1.003 & 1.000 \\ 
  Type-2 & 1.220 & 1.014 & 1.156 & 0.999 & 1.040 & 1.001 & 1.027 & 1.001 & 1.002 & 1.000 \\ 
  Type-3 & 1.298 & 1.150 & 1.223 & 1.020 & 1.053 & 1.001 & 1.026 & 1.001 & 1.002 & 1.000 \\ 
  Type-4 & 1.087 & 1.067 & 1.095 & 1.013 & 1.036 & 1.002 & 1.021 & 1.001 & 1.002 & 1.000 \\ 
  Type-5 & 1.075 & 0.998 & 1.087 & 1.000 & 1.029 & 1.002 & 1.027 & 1.001 & 1.003 & 1.000 \\ 
  Type-6 & 1.011 & 1.000 & 1.023 & 1.002 & 1.016 & 1.002 & 1.006 & 1.001 & 1.002 & 1.000 \\ 
   \bottomrule
\end{tabular}
}
\caption{Comparison of ESA results for various $(N, n)$ pairs and number of 
strong factors in the scenarios with $\Var(\sigma_i^2) = 1$.} 
\label{ESA_table}
\end{table}

Table \ref{ESA_table} gives the average value of each measurement  
over $100$ replications for all of the simulations with mild
heteroscedasticity ($\Var(\sigma^2_i)=1$). 
``Type-1'' to ``Type-6'' correspond to the six cases of factor strengths listed 
in Table \ref{factor_strength_table}.  
The first panel confirms that $m=3$ is broadly effective.
The second panel shows that the problem of PCA is more severe 
at large sample sizes.
The third panel shows 
in contrast that the disadvantage to $m=50$ iterations is more 
severe at the smaller sample sizes.
The fourth panel shows 
similar to the second panel that SVD causes greatest losses at large sample sizes.
The fifth panel shows that ESA has great advantage over QMLE when the variable 
size is large, even at a low aspect ratio $\gamma$.

It remains an interesting puzzle that 
heteroscedasticity is less of a problem when the aspect ratio is higher 
for all the methods. In those
settings there are actually more nuisance $\sigma^2_i$ to estimate. 
One explanation is that no matter what method used, the right factor $R$ of size 
$r \times n$ can be accurately estimated if $\gamma$ is large enough. Then the estimate of 
the left factor $L$ is done via an ordinary linear regression of $Y$ on $R$ which is not 
affected by the heterscedastic noise. This explanation can also work for our observation 
that heteroscedasticity becomes a more severe problem for small $\gamma$, as given $L$, 
it is important to take into consideration different noise variance when estimating $R$.

\subsection*{A.3: Further simulation results}

Here we present more detailed simulation results for the
comparisons among the methods we compare.
All methods used the $m=3$ steps found to be an effective stopping rule.

\begin{table}[h!]
\centering
\scalebox{0.7}{
\begin{tabular}{@{}clcccccccccccc@{}}
  \toprule
  \multirow{2}{1.5cm}{Factor Type} &&\multicolumn{4}{c}{$\gamma =0.02$} & \multicolumn{4}{c}{$\gamma =0.2$} & \multicolumn{4}{c}{$\gamma =1$}\\ 
 \cmidrule(lr){3-6} 
 \cmidrule(lr){7-10} 
 \cmidrule(lr){11-14} 
 &Method & \multicolumn{2}{c}{(20, 1000)} & \multicolumn{2}{c}{(100, 5000)} & \multicolumn{2}{c}{(20, 100)} & \multicolumn{2}{c}{(200, 1000)} & \multicolumn{2}{c}{(50, 50)} & \multicolumn{2}{c}{(500, 500)} \\ 
    \midrule 
 \multirow{2}{1.5cm}{Type-1 \\ 0/6/1/1} & PA & \textbf{0.04} & 5.5 & 0.07 & 7.0 & \textbf{0.12} & 4.9 & 0.10 & 6.9 & \textbf{0.05} & 5.4 & 0.13 & 7.0 \\ 
   & ED & 1.93 & 1.7 & 2.29 & 1.3 & 2.27 & 1.3 & 2.40 & 1.0 & 2.42 & 1.2 & 2.40 & 0.6 \\ 
   & ER & 2.19 & 0.9 & 2.80 & 0.1 & 1.68 & 1.8 & 2.92 & 0.1 & 1.35 & 2.5 & 2.72 & 0.0 \\ 
   & IC1 & 2.30 & 16.0 & 0.69 & 3.3 & 1.44 & 16.0 & 0.61 & 3.5 & 0.10 & 5.6 & 0.69 & 3.1 \\ 
   & NE & 0.23 & 6.3 & 1.82 & 1.3 & 0.16 & 5.0 & 2.45 & 0.6 & 0.08 & 5.4 & 2.36 & 0.5 \\ 
   & BCV & 0.16 & 5.9 & \textbf{0.03} & 5.8 & 0.33 & 4.5 & \textbf{0.01} & 5.9 & 0.12 & 5.0 & \textbf{0.00} & 6.0 \\ 
   & Oracle & -- & 6.0 & -- & 6.0 & -- & 5.9 & -- & 6.0 & -- & 6.0 & -- & 6.0 \\ 
      \midrule 
 \multirow{2}{1.5cm}{Type-2 \\ 2/4/1/1} & PA & 0.27 & 3.7 & 0.15 & 4.6 & 0.55 & 3.4 & 0.34 & 4.0 & 0.69 & 3.2 & 0.31 & 3.9 \\ 
   & ED & 0.61 & 3.5 & 1.03 & 2.9 & 0.95 & 3.0 & 1.18 & 2.5 & 1.00 & 3.0 & 1.03 & 2.6 \\ 
   & ER & 1.52 & 1.8 & 1.21 & 2.0 & 1.64 & 1.9 & 1.33 & 2.0 & 1.34 & 2.0 & 1.23 & 2.0 \\ 
   & IC1 & 1.87 & 16.0 & 0.58 & 3.6 & 1.34 & 16.0 & 0.57 & 3.7 & \textbf{0.09} & 5.8 & 0.66 & 3.2 \\ 
   & NE & 0.42 & 6.6 & 0.87 & 2.7 & \textbf{0.12} & 5.3 & 1.13 & 2.4 & 0.10 & 5.6 & 1.11 & 2.2 \\ 
   & BCV & \textbf{0.26} & 5.4 & \textbf{0.12} & 5.7 & 0.24 & 4.5 & \textbf{0.00} & 5.9 & 0.19 & 4.7 & \textbf{0.00} & 6.0 \\ 
   & Oracle & -- & 5.1 & -- & 5.8 & -- & 5.5 & -- & 6.0 & -- & 5.9 & -- & 6.0 \\ 
      \midrule 
 \multirow{2}{1.5cm}{Type-3 \\ 3/3/1/1} & PA & 0.35 & 3.2 & 0.46 & 3.1 & 0.62 & 3.1 & 0.72 & 3.0 & 0.76 & 3.0 & 0.69 & 3.0 \\ 
   & ED & 0.30 & 4.0 & 0.55 & 4.0 & 0.46 & 3.8 & 0.54 & 3.5 & 0.56 & 3.7 & 0.56 & 3.5 \\ 
   & ER & 4.15 & 1.8 & 16.18 & 2.2 & 3.40 & 1.9 & 13.62 & 2.6 & 0.78 & 3.0 & 0.69 & 3.0 \\ 
   & IC1 & 1.70 & 16.0 & 0.41 & 4.2 & 1.23 & 16.0 & 0.41 & 4.1 & 0.11 & 5.9 & 0.52 & 3.5 \\ 
   & NE & 0.41 & 6.8 & 0.41 & 3.7 & \textbf{0.14} & 5.5 & 0.56 & 3.4 & \textbf{0.10} & 5.6 & 0.60 & 3.2 \\ 
   & BCV & \textbf{0.17} & 5.1 & \textbf{0.26} & 5.3 & 0.26 & 4.5 & \textbf{0.08} & 5.8 & 0.21 & 4.6 & \textbf{0.01} & 5.9 \\ 
   & Oracle & -- & 5.0 & -- & 4.8 & -- & 5.5 & -- & 5.8 & -- & 5.9 & -- & 6.0 \\ 
      \midrule 
 \multirow{2}{1.5cm}{Type-4 \\ 3/1/3/1} & PA & \textbf{0.01} & 3.0 & \textbf{0.02} & 3.0 & \textbf{0.03} & 3.0 & 0.07 & 3.0 & \textbf{0.05} & 3.0 & 0.06 & 3.0 \\ 
   & ED & 0.11 & 3.3 & 0.81 & 4.4 & 0.08 & 3.3 & 0.29 & 3.9 & 0.07 & 3.3 & 0.08 & 3.8 \\ 
   & ER & 5.10 & 1.8 & 19.24 & 2.2 & 3.50 & 1.9 & 16.79 & 2.5 & 3.33 & 2.3 & 0.50 & 3.0 \\ 
   & IC1 & 2.62 & 16.0 & 0.66 & 4.1 & 1.60 & 16.0 & 0.33 & 4.1 & 0.10 & 3.7 & 0.06 & 3.5 \\ 
   & NE & 0.63 & 5.7 & 0.54 & 3.8 & 0.13 & 3.7 & 0.14 & 3.6 & 0.09 & 3.9 & 0.05 & 3.3 \\ 
   & BCV & 0.02 & 3.1 & 0.19 & 3.5 & \textbf{0.03} & 3.3 & \textbf{0.05} & 3.7 & \textbf{0.05} & 3.1 & \textbf{0.01} & 3.9 \\ 
   & Oracle & -- & 3.2 & -- & 3.2 & -- & 3.5 & -- & 3.9 & -- & 3.8 & -- & 4.0 \\ 
      \midrule 
 \multirow{2}{1.5cm}{Type-5 \\ 1/3/3/1} & PA & \textbf{0.02} & 3.4 & \textbf{0.01} & 4.3 & \textbf{0.08} & 3.0 & \textbf{0.01} & 3.8 & 0.10 & 2.9 & 0.02 & 3.7 \\ 
   & ED & 0.40 & 2.0 & 0.58 & 1.9 & 0.54 & 1.6 & 0.56 & 1.6 & 0.57 & 1.6 & 0.45 & 2.0 \\ 
   & ER & 0.69 & 1.0 & 0.78 & 1.0 & 0.70 & 1.0 & 0.79 & 1.0 & 0.71 & 1.0 & 0.72 & 1.0 \\ 
   & IC1 & 2.63 & 16.0 & 0.41 & 2.1 & 1.53 & 16.0 & 0.45 & 2.0 & 0.10 & 3.3 & 0.55 & 1.5 \\ 
   & NE & 0.40 & 5.3 & 0.48 & 1.9 & 0.13 & 3.2 & 0.59 & 1.5 & \textbf{0.08} & 3.5 & 0.62 & 1.2 \\ 
   & BCV & 0.12 & 3.1 & 0.04 & 3.9 & 0.27 & 2.4 & \textbf{0.01} & 3.9 & 0.16 & 2.8 & \textbf{0.00} & 4.0 \\ 
   & Oracle & -- & 3.7 & -- & 4.0 & -- & 4.0 & -- & 4.0 & -- & 4.0 & -- & 4.0 \\ 
      \midrule 
 \multirow{2}{1.5cm}{Type-6 \\ 0/1/6/1} & PA & 0.45 & 5.6 & 0.68 & 7.3 & 0.22 & 4.0 & 2.00 & 10.4 & 0.34 & 4.5 & 2.89 & 12.8 \\ 
   & ED & 0.07 & 0.8 & 0.11 & 1.8 & 0.06 & 0.7 & 0.12 & 1.4 & 0.06 & 0.4 & 0.09 & 1.1 \\ 
   & ER & 0.07 & 0.1 & 0.09 & 0.1 & \textbf{0.03} & 0.2 & 0.08 & 0.1 & 0.05 & 0.1 & 0.06 & 0.1 \\ 
   & IC1 & 3.11 & 13.6 & 0.06 & 1.1 & 1.74 & 16.0 & 0.07 & 1.0 & 0.05 & 0.5 & 0.06 & 0.5 \\ 
   & NE & 0.21 & 3.2 & 0.06 & 1.0 & 0.05 & 0.8 & 0.06 & 0.7 & 0.06 & 0.9 & 0.05 & 0.3 \\ 
   & BCV & \textbf{0.06} & 0.2 & \textbf{0.04} & 1.0 & \textbf{0.03} & 0.1 & \textbf{0.02} & 0.8 & \textbf{0.03} & 0.0 & \textbf{0.00} & 1.0 \\ 
   & Oracle & -- & 1.0 & -- & 1.0 & -- & 0.8 & -- & 1.0 & -- & 0.8 & -- & 1.0 \\ 
   \bottomrule
\end{tabular}
}
\caption{Comparison of REE and $\hat k$ for rank selection methods with
  various $(N, n)$ pairs, and scenarios. For each different scenario, the factors' strengths
  are listed as the number of ``strong/useful/harmful/undetectable'' factors. 
For each $(N, n)$ pair, the first column is the REE and the second column is $\hat k$. Both values are averages over $100$ simulations. 
$\Var(\sigma_i^2) = 1$.
}
\label{compare_k_table}
\end{table}

\begin{table}[ht]
\centering
\scalebox{0.75}{
\begin{tabular}{@{}clcccccccc@{}}
  \toprule
  \multirow{2}{1.5cm}{Factor Type} &&\multicolumn{4}{c}{$\gamma =5$} & \multicolumn{4}{c}{$\gamma =50$}\\ 
 \cmidrule(lr){3-6} 
 \cmidrule(lr){7-10} 
 &Method & \multicolumn{2}{c}{(100, 20)} & \multicolumn{2}{c}{(1000, 200)} & \multicolumn{2}{c}{(1000, 20)} & \multicolumn{2}{c}{(5000, 100)} \\ 
    \midrule 
 \multirow{2}{1.5cm}{Type-1 \\ 0/6/1/1} & PA & \textbf{0.05} & 5.0 & 0.11 & 6.9 & \textbf{0.01} & 5.7 & 0.10 & 7.0 \\ 
   & ED & 1.89 & 1.2 & 1.57 & 1.6 & 0.43 & 4.7 & 0.10 & 6.1 \\ 
   & ER & 2.23 & 0.3 & 2.18 & 0.0 & 1.69 & 0.0 & 1.68 & 0.0 \\ 
   & IC1 & 1.23 & 16.0 & 0.86 & 2.2 & 0.04 & 5.0 & 1.10 & 1.1 \\ 
   & NE & 0.14 & 4.9 & 1.17 & 1.7 & 0.20 & 4.2 & 0.14 & 3.9 \\ 
   & BCV & 0.37 & 4.1 & \textbf{0.00} & 6.0 & 0.10 & 4.9 & \textbf{0.01} & 5.8 \\ 
   & Oracle & -- & 5.9 & -- & 6.0 & -- & 5.8 & -- & 5.9 \\ 
      \midrule 
 \multirow{2}{1.5cm}{Type-2 \\ 2/4/1/1} & PA & 0.68 & 2.8 & 0.23 & 3.9 & 0.32 & 3.1 & 0.12 & 4.0 \\ 
   & ED & 0.83 & 2.9 & 0.65 & 3.2 & 0.17 & 5.2 & 0.06 & 6.0 \\ 
   & ER & 1.05 & 2.0 & 0.94 & 2.0 & 0.95 & 1.9 & 0.68 & 2.0 \\ 
   & IC1 & 1.24 & 16.0 & 0.86 & 2.2 & \textbf{0.05} & 5.0 & 0.68 & 2.0 \\ 
   & NE & \textbf{0.07} & 5.2 & 0.77 & 2.4 & 0.08 & 4.5 & 0.13 & 4.0 \\ 
   & BCV & 0.31 & 4.2 & \textbf{0.00} & 6.0 & 0.09 & 4.9 & \textbf{0.01} & 5.8 \\ 
   & Oracle & -- & 5.9 & -- & 6.0 & -- & 5.7 & -- & 5.9 \\ 
      \midrule 
 \multirow{2}{1.5cm}{Type-3 \\ 3/3/1/1} & PA & 0.59 & 3.0 & 0.51 & 3.0 & 0.35 & 3.0 & 0.35 & 3.0 \\ 
   & ED & 0.48 & 3.6 & 0.36 & 3.9 & 0.11 & 5.5 & 0.06 & 6.2 \\ 
   & ER & 3.51 & 1.9 & 22.02 & 2.1 & 3.33 & 2.0 & 15.40 & 2.0 \\ 
   & IC1 & 1.27 & 16.0 & 0.48 & 3.1 & \textbf{0.04} & 5.0 & 0.35 & 3.0 \\ 
   & NE & \textbf{0.09} & 5.2 & 0.47 & 3.1 & 0.05 & 4.7 & 0.14 & 3.9 \\ 
   & BCV & 0.25 & 4.5 & \textbf{0.01} & 5.8 & 0.09 & 4.6 & \textbf{0.01} & 5.8 \\ 
   & Oracle & -- & 5.9 & -- & 6.0 & -- & 5.8 & -- & 5.9 \\ 
      \midrule 
 \multirow{2}{1.5cm}{Type-4 \\ 3/1/3/1} & PA & \textbf{0.03} & 3.0 & 0.03 & 3.0 & \textbf{0.01} & 3.0 & \textbf{0.01} & 3.0 \\ 
   & ED & 0.05 & 3.2 & 0.05 & 3.6 & \textbf{0.01} & 3.3 & 0.03 & 4.0 \\ 
   & ER & 3.36 & 1.8 & 25.02 & 2.1 & 3.67 & 2.0 & 18.55 & 2.0 \\ 
   & IC1 & 1.53 & 16.0 & 0.03 & 3.1 & \textbf{0.01} & 3.0 & \textbf{0.01} & 3.0 \\ 
   & NE & 0.04 & 3.4 & 0.03 & 3.2 & \textbf{0.01} & 3.0 & \textbf{0.01} & 3.0 \\ 
   & BCV & \textbf{0.03} & 3.2 & \textbf{0.01} & 3.8 & \textbf{0.01} & 3.2 & \textbf{0.01} & 3.7 \\ 
   & Oracle & -- & 3.8 & -- & 4.0 & -- & 3.6 & -- & 3.8 \\ 
      \midrule 
 \multirow{2}{1.5cm}{Type-5 \\ 1/3/3/1} & PA & \textbf{0.11} & 2.7 & \textbf{0.01} & 3.6 & \textbf{0.01} & 3.1 & \textbf{0.00} & 4.0 \\ 
   & ED & 0.42 & 1.8 & 0.32 & 2.1 & 0.31 & 1.9 & 0.12 & 3.7 \\ 
   & ER & 0.57 & 1.0 & 0.57 & 1.0 & 0.43 & 1.0 & 0.42 & 1.0 \\ 
   & IC1 & 1.45 & 16.0 & 0.54 & 1.1 & 0.34 & 1.3 & 0.42 & 1.0 \\ 
   & NE & 0.12 & 2.8 & 0.53 & 1.1 & 0.08 & 2.5 & 0.15 & 2.0 \\ 
   & BCV & 0.22 & 2.4 & \textbf{0.01} & 3.9 & 0.12 & 2.6 & 0.02 & 3.8 \\ 
   & Oracle & -- & 3.9 & -- & 4.0 & -- & 3.7 & -- & 3.8 \\ 
      \midrule 
 \multirow{2}{1.5cm}{Type-6 \\ 0/1/6/1} & PA & 0.29 & 3.4 & 2.27 & 10.5 & 0.77 & 5.4 & 1.24 & 7.1 \\ 
   & ED & 0.03 & 0.2 & 0.04 & 0.6 & 0.02 & 0.5 & 0.03 & 0.9 \\ 
   & ER & \textbf{0.02} & 0.0 & 0.04 & 0.0 & \textbf{0.01} & 0.0 & \textbf{0.01} & 0.0 \\ 
   & IC1 & 1.00 & 7.4 & 0.03 & 0.1 & \textbf{0.01} & 0.0 & \textbf{0.01} & 0.0 \\ 
   & NE & 0.03 & 0.2 & 0.03 & 0.2 & \textbf{0.01} & 0.0 & \textbf{0.01} & 0.0 \\ 
   & BCV & \textbf{0.02} & 0.1 & \textbf{0.01} & 0.8 & \textbf{0.01} & 0.1 & 0.02 & 0.7 \\ 
   & Oracle & -- & 0.5 & -- & 0.9 & -- & 0.6 & -- & 0.8 \\ 
   \bottomrule
\end{tabular}
}
\caption{Like Table \ref{compare_k_table}, but for larger $\gamma$.}
\label{compare_k_table_con}
\end{table}

\end{document}